\documentclass[a4paper,12pt]{article}
\usepackage[utf8x]{inputenc}
\usepackage[left=2cm,right=2cm,top=2cm,bottom=2cm]{geometry}

\usepackage[linktoc=all,hidelinks]{hyperref}
\usepackage{cite}
\usepackage{calrsfs}
\usepackage{amsmath}
\usepackage{amsfonts}
\usepackage{amssymb}
\usepackage[normalem]{ulem}

\usepackage{tensor}
\usepackage[smalltableaux,centertableaux,centerboxes]{ytableau}

\newcommand{\pd}{\partial}
\newcommand{\cL}{\mathcal{L}}

\newcommand{\cK}{\mathcal{K}}
\newcommand{\cR}{\mathcal{R}}

\newcommand{\overbar}[1]{\mkern 2mu\overline{\mkern-2mu#1}}

\renewcommand*{\bar}{\overbar}
\newcommand{\bbar}[1]{\bar{\bar{#1}}}
\newcommand{\bbbar}[1]{\bar{\bar{\bar{#1}}}}

\begin{document}
\numberwithin{equation}{section}

\thispagestyle{empty}
\begin{center}

\vspace*{50pt}
{\LARGE \bf
Higher spins from exotic dualisations}

\vspace{30pt}
{Nicolas Boulanger${}^{\,a}$ and Victor Lekeu${}^{\, b}$}

\vspace{10pt}
\texttt{nicolas.boulanger@umons.ac.be, victor.lekeu@aei.mpg.de}

\vspace{20pt}
\begin{enumerate}
\item[${}^a$] {\sl \small
Physique de l’Univers, Champs et Gravitation, Université de Mons - UMONS,\\
Place du Parc 20, 7000 Mons, Belgium}
\item[${}^b$] {\sl \small
Max-Planck-Institut für Gravitationsphysik (Albert-Einstein-Institut)\\
Am Mühlenberg 1, 14476 Potsdam, Germany}
\end{enumerate}

\vspace{50pt}
{\bf Abstract} 
\end{center}

\noindent

At the free level, a given massless field can be described by an infinite number of different potentials related to each other by dualities. In terms of Young tableaux, dualities replace any number of columns of height $h_i$ by columns of height $D-2-h_i$, where $D$ is the spacetime dimension: in particular, applying this operation to empty columns gives rise to potentials containing an arbitrary number of groups of $D-2$ extra antisymmetric indices. Using the method of parent actions, action principles including these potentials, but also extra fields, can be derived from the usual ones. In this paper, we revisit this off-shell duality and clarify the counting of degrees of freedom and the role of the extra fields. Among others, we consider the examples of the double dual graviton in $D=5$ and two cases, one topological and one dynamical, of exotic dualities leading to spin three fields in $D=3$.

\newpage

\setcounter{tocdepth}{2}
\tableofcontents

\newpage


\section{Introduction}

Electric-magnetic duality for linearised gravity in Minkowski 
spacetime has received a renewed interest since the original works 
\cite{Hull:2000zn,Hull:2000rr,Hull:2001iu,West:2001as,West:2002jj}. 
The dual graviton in space-time dimension $D$ is given by a gauge field
\begin{equation}\label{eq:Cantisym}
    C_{a_1\ldots a_{D-3}|b} = C_{[a_1\ldots a_{D-3}]|b} 
\end{equation}
of $GL(D)$-irreducible symmetry type $[D-3,1]\,$, that is, obeying the over-antisymmetrisation constraint 
\begin{equation}
    C_{[a_1\ldots a_{D-3}|b]} = 0
\end{equation}
on top of \eqref{eq:Cantisym}. The Bianchi identities of the usual graviton are 
equivalent to the equations of motion for the dual graviton field, 
and vice-versa \cite{Hull:2001iu}. 
We call this type of relation \emph{on-shell duality}. 
In the papers \cite{West:2001as,West:2002jj}, this was 
extended \emph{off-shell}: covariant field equations for the 
dual graviton around eleven-dimensional Minkowski spacetime 
were derived from the Einstein-Hilbert action.
The action principle for the $[D-3,1]$-type gauge field 
$C_{a_1\ldots a_{D-3}|b}$ obtained by the off-shell dualisation 
procedure of \cite{West:2001as} was subsequently spelled out 
in \cite{Boulanger:2003vs}. Taken in flat spacetime of 
dimension $D=5\,$, that action reproduces the one studied long 
ago by Curtright in \cite{Curtright:1980yk}, and in general 
dimension $D\,$, it reproduces the action given in 
\cite{Aulakh:1986cb}.

On-shell duality can be generalised in several different ways. 
Already for the graviton, since the Riemann tensor has two 
pairs of antisymmetric indices, one could consider dualising 
twice, once on each pair. This leads to a field 
$D_{a_1\ldots a_{D-3}|b_1\ldots b_{D-3}}$ with 
$GL(D)$-irreducible symmetry type $[D-3,D-3]\,$, called the 
double-dual graviton \cite{Hull:2001iu}. 
One could also dualise on one or more empty set of indices; 
this leads to an infinite number of equivalent descriptions of 
linearised gravity with extra sets of $D-2$ antisymmetric 
indices which were already mentioned by Siegel to the author of \cite{Hull:2001iu}, 
in a private communication.
Their $GL(D)$-irreducible 
symmetry types are $[D-2,\ldots,D-2,1,1]$ for the higher duals of 
the Fierz-Pauli field $h_{ab}$, 
$[D-2,\ldots,D-2,D-3,1]$ for the higher duals of the dual 
graviton
$C_{a_1 \dots a_{D-3}|b}$ and $[D-2,\ldots,D-2,D-3,D-3]$ for 
the higher duals of the double-dual graviton 
$D_{a_1\ldots a_{D-3}|b_1\ldots b_{D-3}}$.
In terms of Young tableaux, the $GL(D)$ symmetry types of 
these 
fields are obtained by sticking to the left of the Young tableaux of the three 
original fields an arbitrary number of columns of height $D-2\,$. Remarkably, it 
appears that all the $GL(11)$-irreducible representations corresponding to the 
infinite tower of fields associated with the dual graviton $C_{a_1 \dots a_8|b}\,$ 
are present in the decomposition of the adjoint 
representation of $e_{11}\,$ \cite{Riccioni:2006az}.

More generally, consider 
mixed-symmetry gauge fields of arbitrary symmetry type. 
(This includes in particular the case of higher-spin fields, for which each 
column has one box). 
Then, one could dualise on any number of columns of height $h_i$, replacing them with 
columns of height 
$D-2-h_i$ \cite{Hull:2001iu,Bekaert:2002dt,deMedeiros:2002qpr,Bekaert:2003az}, 
or dualise on empty columns to add an arbitrary number of columns of height 
$D-2$ to the left of the Young diagram \cite{Hull:2001iu,Riccioni:2006az,Boulanger:2012df,Boulanger:2015mka}. 

In maximal supergravity theories, this collection of mixed symmetry fields arising 
from higher dualisations, also called \emph{exotic} dualisations,  of the usual potentials plays a crucial role and have been 
conjectured to couple to the various `exotic branes' of string theory 
\cite{West:2004kb,Cook:2009ri,Bergshoeff:2011se,Chatzistavrakidis:2013jqa,West:2018lfn,Fernandez-Melgarejo:2018yxq}. Notice however that each 
successive dualisation exchanges the role of Bianchi identities and equations of 
motion: therefore, an even number of dualisations does \emph{not} exchange them. 
Indeed, at least at the linear and classical level there are only two types of 
sources, electric and magnetic \cite{Hull:2001iu}, and the potentials for an even 
number of dualisations are related locally up to a gauge transformation, 
as has been recently pointed out in \cite{Henneaux:2019zod,Chatzistavrakidis:2019bxo} (see however \cite{Chatzistavrakidis:2020kpx}). 
This is to be contrasted with the non-local relation between a $p$-form 
and its dual $D-p-2$-form, or between the Fierz-Pauli field $h_{ab}$ and 
the first dual $C_{a_1 \dots a_{D-3}|b}$ (more generally, between potentials 
related by an odd number of dualisations).

The above discussion concerned various aspects of \emph{on-shell} duality; 
instead, this paper deals with \emph{off-shell} dualities. 
Difficulties in finding a manifestly Poincar\'e-covariant action principle 
for the double dual graviton were mentioned already in \cite{Hull:2001iu}, 
since the tensorial structure of the left-hand side of the field equations 
does not coincide with the one of the gauge field itself.
For example, in dimension $D=5\,$, the field equations for the double-dual
graviton $D_{ab,cd}$ are the double-trace equations
\begin{align}
  \bbar{K}_{ad}:=\eta^{be}\eta^{cf}K_{abc|def}=0\;, 
\end{align}
where
\begin{equation}
    K_{abc|}{}^{def}:=9\,\partial_{[a}\partial^{[d}D_{bc]|}{}^{ef]}
\end{equation}
is the gauge-invariant field strength for the double-dual graviton.

Following these works and motivated by the $e_{11}$ proposal of West, 
a manifestly Poincar\'e-invariant action was proposed in the paper \cite{Boulanger:2012df} for the double-dual graviton, obtained by 
extending the off-shell dualisation procedure of \cite{West:2001as,Boulanger:2003vs}.
In fact, not only was an algorithm proposed for the double dualisation of 
the Fierz-Pauli field, but also for all the actions featuring a member of 
the three infinite towers of higher duals of the gravitons. 
A general procedure to build the parent actions for arbitrary higher-dual fields 
was later given in \cite{Boulanger:2012mq} and applied to several cases 
including double field theory in \cite{Boulanger:2015mka,Bergshoeff:2016ncb,Bergshoeff:2016gub,Chatzistavrakidis:2019len}. 
For higher duals, this procedure produces actions that contain some extra fields 
that are necessary to reproduce the correct higher trace equations 
of motion and counting of degrees of freedom.
The main purpose of this paper is to clarify this latest point 
and the role of the extra fields. 

Let also us mention that different actions for the double-dual graviton 
and other higher duals have appeared recently in the note \cite{Henneaux:2019zod}. 
They exploit the local \emph{on-shell} relation mentioned above 
between potentials that arise from an even number of dualisations. 
In the example of the double-dual graviton in $D=5\,$, 
this on-shell relation reads
\begin{equation}
    D^{ab|}{}_{cd} = \delta^{[a}{}_{[c} h^{b]}{}_{d]} 
    + \mathbb{P}_{[2,2]}(\pd_c \xi^{ab|}{}_d )\;,
\end{equation}
where $\xi^{ab|}{}_c$ is the $[2,1]$-gauge parameter of the double-dual 
graviton and $\mathbb{P}_{[2,2]}$ denotes the projector onto the $[2,2]$ Young 
symmetry (so that the second term is just the usual gauge transformation of 
$D_{ab|cd}$). This relation can be inverted to obtain 
$h_{ab} = H_{ab}(D,\pd \xi)$ as a function of $D$ and $\xi\,$.
Then, plugging this expression into the usual Fierz-Pauli Lagrangian gives an action 
$S[D,\xi] = S_\text{FP}[H(D,\pd\xi)]$ which reproduces the correct double trace 
equation of motion for $D$ by construction. The traceless part of $D$ does 
not appear in that action and $\xi$ can be gauged away by a shift symmetry; 
doing so reproduces the Fierz-Pauli action in its standard form. 
This is in sharp contrast from what happens in the actions 
considered in \cite{Boulanger:2012df} and the present 
paper: in particular, while they both contain some extra 
fields, here these fields \emph{cannot} be gauged away and 
the action for the double dual graviton is not a rewriting 
of the Fierz-Pauli action. The mechanism leading to the 
propagation of the correct degrees of freedom is different, 
as we clarify in the present paper.

This paper is organized as follows:
\begin{itemize}
    \item In section \ref{sec:twocolumns}, we first consider representative cases 
    of gauge fields associated with Young tableaux with at most two columns. 
    The actions then only involve the higher dual field and the original field, 
    both with their usual gauge symmetries. There is no other extra field off-shell. 
    Both fields appear with the standard kinetic terms, one of which carries the 
    wrong sign, along with a gauge-invariant 
    cross-term. The action reproduces the duality relation between the two fields, 
    which enforces the correct counting of degrees of freedom and the higher trace equations of motion. This is done through 
    the following examples, which are sufficient to convince oneself of the general 
    case:
    \begin{enumerate}
        \item scalar -- graviton in $D=3$;
        \item vector -- Curtright in $D=4$; and
        \item graviton -- `window' in $D=5$.
    \end{enumerate}
    Here and in the rest of the paper, a Curtright (or `hook') field refers to a $[2,1] \sim \ydiagram{2,1}$ mixed symmetry field and `window' to a field of type 
    $[2,2] \sim \ydiagram{2,2}\,$.
    \item In section three, we consider two cases with three columns:
    \begin{enumerate}
        \item graviton - spin three in $D=3$; and
        \item vector - spin three in $D=3$.
    \end{enumerate}
    The first example studied is topological, 
    while the second one leads to an interesting action containing a spin three 
    field propagating one degree of freedom. 
    Here, the result of the two-column case cannot hold: indeed, 
    it is not possible to write down a two-derivative, gauge-invariant 
    cross-terms between those fields with the usual gauge transformations. 
    Instead, we find that the action also involves more fields with less 
    columns and their own gauge invariances. 
    Furthermore, the gauge variations of the various fields 
    cannot be disentangled, 
    i.e., every given field will also transform with the gauge parameters 
    of the other fields, whenever Lorentz covariance makes it possible. 
    As a consequence, when computing the Euler-Lagrange equations of motion
    we find that the method of parent actions 
    gives a version of the on-shell duality relations  
    that is invariant under these entangled gauge symmetries.
    These features are expected to hold in more general cases.
\end{itemize}

\paragraph{Conventions.} Square brackets denote complete antisymmetrisation and parentheses complete symmetrisation, both with strength one. We are in flat Minkowski space throughout, with mostly plus signature, and use Latin indices. Traces of tensors are denoted by a bar except when no confusion can arise, in which case it is only indicated by the lower number of indices.

\section{The two column case}
\label{sec:twocolumns}

\subsection{Scalar -- graviton in \texorpdfstring{$D=3$}{D=3}}
\label{sec:3Dscalarmetric}

We first consider the simplest case of a massless scalar field $\phi$ in three dimensions. Seen as a degenerate two-column Young tableau with empty columns, its double dual is a symmetric field $h_{ab}\,$.

\paragraph{On-shell dualisation and equations of motion.}

In this context, the `curvature' of the scalar field $\phi$ is the symmetric tensor
\begin{equation}
    K_{mn}[\phi] := \pd_m \pd_n \phi \sim \ydiagram{2}\, .
\end{equation}
The equation of motion for $\phi$ is the Klein-Gordon equation $\pd_m \pd^m \phi = 0$, which in this language is the tracelessness of the curvature,
\begin{equation}
    \bar{K}[\phi] := K\indices{^m_m}[\phi] = 0\, .
\end{equation}

Dualising twice (once on each column), we then define the tensor
\begin{equation}\label{eq:doubledual}
    \cR\indices{^{ab}_{cd}} := \varepsilon^{abm} \varepsilon_{cdn} K\indices{_m^n} = 4 \,\delta^{[a}_{[c} K\indices{^{b]}_{d]}} - \delta^{ab}_{cd} \bar{K} \sim \ydiagram{2,2}\, ,
\end{equation}
with inverse relation
\begin{equation}
    K_{mn} = \frac{1}{4}\,\varepsilon_{mab} \varepsilon_{ncd} \cR^{abcd} = \bar{\cR}_{mn} - \frac{1}{2} \eta_{mn} \bbar{\cR}\, .
\end{equation}
On-shell, we have $\pd^m K_{mn} = 0$, which implies 
the Bianchi identity $\pd_{[m} \cR_{np]qr} = 0\,$. 
This then guarantees the existence of a symmetric tensor $h_{ab}$ 
such that $\cR$ is the (linearized) Riemann tensor of $h_{ab}$,
\begin{equation}
\label{eq:defR}
\cR\indices{^{ab}_{cd}} = 4\, \pd^{[a}\pd_{[c} 
h\indices{^{b]}_{d]}}\, ,
\end{equation}
and where we stress that the equality only holds when the massless Klein-Gordon
equation is satisfied. 
The field $h_{ab}$ is the double dual of $\phi\,$ \cite{Hull:2001iu}. 
Its equation of motion is the \emph{double} trace equation
\begin{equation}\label{eq:rbarbar}
    \bar{\bar{\cR}}[h]:= \cR\indices{^{ab}_{ab}}[h]=0\, ,
\end{equation}
as follows from the duality relations \eqref{eq:doubledual}. Notice that this is invariant under the usual gauge transformations of linearised gravity,
\begin{equation}
    \delta h_{ab} = 2 \,\pd_{(a}\epsilon_{b)}\, .
\end{equation}
Nevertheless, $h_{ab}$ propagates one degree of freedom and not zero in three 
dimensions, since its equation of motion \eqref{eq:rbarbar} is weaker than the 
(linearised) vacuum Einstein equations $\bar{\cR}_{ab} = 0$.

On-shell, this duality relation becomes 
$\bar{\cR}_{ab} = K_{ab} \equiv \pd_a \pd_b \phi\,$ or, equivalently,
\begin{equation}
    \cR\indices{^{ab}_{cd}} =4\, \pd^{[a}\pd_{[c} h\indices{^{b]}_{d]}} = 4\, \pd^{[a}\pd_{[c} \delta^{b]}_{d]} \phi \, .
\end{equation}
This implies that $\phi$ and $h_{mn}$ are related locally and algebraically (up to a gauge transformation) as
\begin{equation}
h_{mn} = \eta_{mn} \,\phi + 2 \,\pd_{(m} \xi_{n)}\, .
\end{equation}
This is a general phenomenon when there are an even number of dualisations \cite{Henneaux:2019zod,Chatzistavrakidis:2019bxo}.

\paragraph{Action principle.}

The above relations follow from the action
\begin{align}
    S[\phi,h_{ab}]  &= 
    \int d^3x \left(\cL_\text{FP} + \tfrac{1}{2}\,\partial_a \phi\,\partial^a \phi  
+ \tfrac{1}{2} \bbar{\cR}[h] \phi\right) \label{eq:laghphi3D}\\
&= 
    \int d^3x \big[ \cL_\text{FP} + \tfrac{1}{2}\,\partial_a \phi\,\partial^a \phi  
+ \partial_a\phi \,(\partial_b h^{ab} - \partial^a h)\big]\;,
\end{align}
(with $h=h^a{}_a$) where $\cL_\text{FP}$ is the usual Fierz-Pauli Lagrangian for $h_{ab}$,
\begin{equation}\label{eq:fierzpauli}
    \cL_\text{FP} = -\tfrac{1}{2}\,\partial_a h_{bc}\,\partial^a h^{bc}\,
    +\tfrac{1}{2}\,\partial_a h\,\partial^a h\,
    - \partial_a h^{ab}\,\partial_b h + 
    \partial_a h^{ab}\,\partial^c h_{cb}\, .
\end{equation}
Note the wrong sign kinetic term for $\phi$. The action is invariant under the usual gauge symmetries
\begin{align}
    \delta h_{ab} &= 2 \,\pd_{(a} \epsilon_{b)} \qquad \delta \phi = 0\; .
\end{align}
This is the main result of this section; it is derived from the action principles of \cite{Boulanger:2012df,Boulanger:2015mka} after a few field redefinitions and extra dualisations that are explained below.

The equations of motions coming from this Lagrangian are
\begin{align}
    - \Box \phi + \frac{1}{2} \bbar{\cR}[h] &= 0 \\
    \bar{\cR}_{mn}[h] - \frac{1}{2} \eta_{mn} \bbar{\cR}[h] - 
    \left( \pd_m \pd_n \phi  - \eta_{mn} \Box \phi\right) &= 0 \, . \label{eq:tensoreq3D}
\end{align}
Taking the trace of the second one, we get 
$+ 2 \Box \phi - \frac{1}{2} \bbar{\cR}[h] = 0$; 
linear combinations of this with the first equation gives
\begin{align}
\Box \phi &= 0 \\
\bbar{\cR}[h] &= 0\, .
\end{align}
In particular, the double trace equation for $h_{ab}$ comes naturally 
out of this Lagrangian. Moreover, using these relations in the tensor equation \eqref{eq:tensoreq3D} gives the duality relation
\begin{equation}
    \bar{\cR}_{mn}[h] = \pd_m \pd_n \phi
\end{equation}
between $h_{mn}$ and $\phi$, which ensures that there 
is no extra degree of freedom.

In order to obtain this action, we start from the action (2.3.5) of 
\cite{Boulanger:2015mka}, 
\begin{align}
    S[Y^{ab|}{}_c] = \int d^D\! x \,\big(
    \tfrac{1}{2}\,\partial_c Y^{ca|}{}_b\partial^d Y_{da|}{}^b\,
    -\tfrac{1}{2(D-1)}\,\partial_c Y^{ca|}{}_a\partial^b Y_{bd|}{}^d
    \big)\label{eq:actionY}
\end{align}
that features the field $Y\indices{^{ab|}_c}\,$ antisymmetric in 
its two upper indices but with no extra conditions otherwise. 
That action was obtained \cite{Boulanger:2015mka} by dualising Maxwell's action, 
which is itself the dual of a scalar in $3D\,$, so this is indeed what we should 
look at if one wants to double dualise the massless scalar.
In order to be self-contained, let us recall from \cite{Boulanger:2015mka} how 
this procedure works. Up to integration by parts, 
the Maxwell action can be written in the form

\begin{align}\label{eq:maxP}
    S[A_a] &= \frac{1}{2}\int d^D\!x\left( - \pd_a A_b \,\pd^a A^b + \pd_a A^a \,\pd_b A^b \right) \\
    &= \frac{1}{2}\int d^D\!x\left( - P_{ab}[A] P^{ab}[A] + P\indices{_a^a}[A]P\indices{_b^b}[A] \right)\, , \quad P_{ab}[A] := \pd_a A_b\, ,
\end{align}
where $P_{ab}[A]$ contains both a symmetric and an antisymmetric part, unlike the usual $F_{ab}[A] = 2\pd_{[a} A_{b]}$. One then introduces the parent action
\begin{equation}\label{eq:parentmax}
    S[P_{ab},Y\indices{^{ab|}_c}] = \int d^D\!x\left( - \tfrac{1}{2} P_{ab}\, P^{ab} + \tfrac{1}{2} P\indices{_a^a} P\indices{_b^b} + P\indices{_a^b}\, \pd_c Y\indices{^{ca|}_b} \right)\, .
\end{equation}
The equation of motion for the $Y\indices{^{ab|}_c}$ field is $ \pd_{[a} P\indices{_{b]}^c} = 0$ , which implies (using the Poincaré lemma) that $P_{ab} = \pd_a A_b$ for some vector field $A_a$. Plugging this solution back into \eqref{eq:parentmax} reproduces the Maxwell action \eqref{eq:maxP}. On the other hand, $P_{ab}$ is an auxiliary field in \eqref{eq:parentmax}: its equation of motion, which reads $- P_{ab} + \eta_{ab} P\indices{_c^c} + \pd^c Y_{ca|b} = 0$,
can be solved algebraically as \begin{equation}
    P_{ab} = \pd^c Y_{ca|b} - \tfrac{1}{(D-1)}\, \eta_{ab} \pd_c Y\indices{^{cd|}_d}\, .
\end{equation}
Plugging this expression into \eqref{eq:parentmax} leads to the dual action \eqref{eq:actionY}.

The field $Y\indices{^{ab|}_c}$ is decomposed as
\begin{equation}
    Y^{ab|}{}_{c} = X\indices{^{ab|}_c} 
    + 2\,\delta^{[a}_{c} Z^{b]}\; ,\quad X\indices{^{ab}_b} = 0\;,
\end{equation}
where the three-index tensor $X\indices{^{ab|}_c}$ is traceless \cite{Boulanger:2015mka}. 
In dimension $D=3\,$, one can trade a pair of antisymmetric indices for a single one: the field $Y^{ab|}{}_c\,$ then decomposes 
according to 
\begin{equation}\label{eq:Ydecomp3D}
    Y^{ab|}{}_{c} = \varepsilon^{abd} h_{dc} 
    + 2\,\delta^{[a}_{c} Z^{b]}\; ,\quad h_{ab}=h_{ba}\;,
\end{equation}
where $h_{ab}$ is symmetric so that the first piece in the 
decomposition of $Y\indices{^{ab|}_c}$ is traceless. 
This decomposition gives the following action 
\begin{align}
    S[h_{ab},Z_a] =& \int d^3x \left(
    -\tfrac{1}{2}\,\partial_{a}h_{bc}\,\partial^{a}h^{bc}
    + \tfrac{1}{2}\,\partial_{a}h_{bc}\,\partial^{b}h^{ac}
    + \tfrac{1}{2}\,\varepsilon^{bcd}\,\partial^a h_{ab}\,F_{cd}[Z]
    + \tfrac{1}{4}\,F^{ab}[Z]F_{ab}[Z] \right)\;,
    \label{dualspin1spin2}
\end{align}
where $F_{ab}(Z) = 2\,\partial_{[a}Z_{b]}\,$. This action is invariant under
\begin{equation}
    \delta h_{ab} = 2\, \pd_{(a} \epsilon_{b)} \, , \quad 
    \delta Z_a = \pd_a \lambda + \varepsilon_{abc} \,
    \pd^b \epsilon^c\;.
    \label{gaugetransfodualspin1spin2}
\end{equation}
Notice the mixing of gauge symmetries: in particular, the pure $h_{mn}$ 
part of the Lagrangian is \emph{not} Fierz-Pauli, but its non 
gauge invariance is compensated by the cross-term between $h$ and $Z\,$.

Crucially, because of the $\lambda$ gauge invariance, 
this action depends (up to boundary terms) on $Z_a$ only 
through its field strength $F_{ab} = 2\, \pd_{[a} Z_{b]}\,$.
This makes it possible to dualise $Z_a$ to a scalar $\varphi$ through 
the usual procedure, i.e., by adding the term 
$-\varphi \varepsilon^{abc} \pd_{[a} F_{bc]}$ 
to the Lagrangian \eqref{dualspin1spin2} where $F=F[Z]$ is not imposed. 
Extremising the resulting action with respect to $\varphi$ reproduces $F = F[Z]$ 
and  the previous Lagrangian, while eliminating the auxiliary field $F$ 
produces the dual action 
\begin{align}
    S[\varphi,h_{ab}]  = 
    \int \!d^3\!x\, \Big[ 
-\tfrac{1}{2}\,\partial_a h_{bc}\,\partial^a h^{bc}
+\,\partial_a h^{ab}\,\partial^c h_{bc}\, 
+ 2\,\partial_a\varphi \,(\partial^a\varphi + \partial_b h^{ab}) 
    \Big]
\end{align}
that is invariant under
\begin{equation}
    \delta h_{ab} = 2 \,\pd_{(a} \epsilon_{b)} \, , \quad 
    \delta \varphi = - \pd^a \epsilon_a \, .
\end{equation}
This gauge transformation suggests the simple change of variable 
$\phi := 2\varphi + \eta^{ab}h_{ab}\,$ 
in terms of a gauge-invariant scalar field $\phi\,$. 
This gives the dual action $S[\phi,h_{ab}]$ 
introduced at the start of the section:
\begin{align}
    S[\phi,h_{ab}]  = 
    \int d^3x \big[& 
-\tfrac{1}{2}\,\partial_a h_{bc}\,\partial^a h^{bc}
+ \tfrac{1}{2}\,\partial_a h\,\partial^a h
- \partial_a h\,\partial_b h^{ab} 
+\,\partial_a h^{ab}\,\partial^c h_{bc}\, 
\nonumber \\
& + \tfrac{1}{2}\,\partial_a \phi\,\partial^a \phi  
+ \partial_a\phi \,(\partial_b h^{ab} - \partial^a h)\big] \;.
\label{dualscalarh}
\end{align}

\subsection{Vector -- hook in \texorpdfstring{$D=4$}{D=4}}

In a similar way, the double dual of a Maxwell field $A_a$ (seen as a two-column, $[1,0]$ tensor) in four spacetime dimensions is a $[2,1]$ Curtright field $T_{ab|c}$, i.e. satisfying
\begin{equation}
    T_{ab |c} = - T_{ba|c} \, , \qquad T_{[ab|c]} = 0\, .
\end{equation}
Equivalently, this can be seen as a single `exotic dualisation', adding a column of height $D-2 = 2$, of the field $\tilde{A}_a$ which is the conventional dual of $A_a$.

The equations of motion and duality relations are obtained from the action
\begin{equation}\label{eq:lagAT4D}
    S[T_{ab|c},A_a] = \int \! d^4\!x \left( \cL_\text{C} + \tfrac{1}{4}\, F_{mn}[A] F^{mn}[A] 
    - \tfrac{1}{\sqrt{2}}\, A_a \bar{\bar{K}}{}^a[T] \right) \, .
\end{equation}
Here, $\cL_\text{C}$ is the Curtright Lagrangian \cite{Curtright:1980yk} for the mixed symmetry field $T_{ab|c}$,
\begin{equation}\label{eq:curtrightlag}
    \cL_\text{C} = -\tfrac{1}{6}\, 
\left( F^{abc|d}[T]\,F_{abc|d}[T]
-3\, F^{abc|}{}_{c}[T]\,F_{abd|}{}^{d}[T] \right)
\end{equation}
where
\begin{equation}
    F^{abc|d}[T] = 3\,\pd^{[a} T^{bc]|d}\, .
\end{equation}
The field strength of $A_a$ is the usual $F_{mn}[A] = 2\pd_{[m} A_{n]}$, the curvature of $T_{ab|c}$ is defined as
\begin{equation}
    K^{abc|}{}_{mn}[T] = 6\, \pd^{[a} \pd_{[m} T^{bc]|}{}_{n]}
\end{equation}
and its traces as $\bar{K}{}^{ab|}{}_c = K\indices{^{abd|}_{cd}}\,$, 
$\bar{\bar{K}}{}^a = K\indices{^{abc|}_{bc}}$.
All in all, the action \eqref{eq:lagAT4D} has the same structure as the action \eqref{eq:laghphi3D} of the last section: it is the sum of the Curtright action, a wrong-sign Maxwell action, and a gauge-invariant cross-term. The action is invariant under the usual independent gauge symmetries of each field,
\begin{align}
    \delta A_a &= \pd_a \lambda\, , \label{gaugeA4D}\\
    \delta T_{abc} &= 2 \,\pd_{[a} a_{b]c} + 2 \,\pd_{[a} s_{b]c} - 2 \,\pd_c a_{ab} \qquad (a_{ab}=-a_{ba}\,, \; s_{ab} = s_{ba})\, . \label{gaugeT4D}
\end{align}
The cross-term is invariant by virtue of the contracted Bianchi identity
\begin{equation}
    \pd_a \bar{\bar{K}}{}^a[T]=0
\end{equation}
and the invariance of $K^{abc|}{}_{mn}[T]$ itself under the gauge symmetries of $T_{ab|c}$. (Note that $F^{abc|d}[T]$ is by itself only invariant under the $s_{ab}$ gauge symmetry; however, the combination \eqref{eq:curtrightlag} is also invariant under the $a_{ab}$ gauge symmetry \cite{Curtright:1980yk}.)

The equations of motion are then
\begin{align}
    \pd_b F^{ab} + \tfrac{1}{\sqrt{2}} \bar{\bar{K}}{}^a &= 0\\
    -\bar{K}{}\indices{^{ab|}_c} - \delta^{[a}_c \bar{\bar{K}}{}^{b]} + \tfrac{1}{\sqrt{2}}\, \pd_{c} F^{ab} + \tfrac{1}{\sqrt{2}}\, \delta^{[a}_c \pd_d F^{b]d} &= 0\, .\label{eq:eomG}
\end{align}
Taking the trace of the second one and combining it with the first, one finds the required equations of motion
\begin{align}
\pd_a F^{ab} &= 0 \\
\bbar{K}{}^a &= 0\, .\label{eq:ttG}
\end{align}
Using these in \eqref{eq:eomG}, this yields the duality relation between the two fields,
\begin{equation}\label{eq:dualityAT}
    \bar{K}{}\indices{^{ab|}_c} = \tfrac{1}{\sqrt{2}} \,\pd_{c} F^{ab}\, .
\end{equation}
Indeed, the duality relation can be written as (with a choice of normalisation)
\begin{align}
    \tfrac{1}{\sqrt{2}} \pd_m F^{ab} &= -\tfrac{1}{12}\, \varepsilon_{mpqr} \,\varepsilon^{abcd}\, K\indices{^{pqr|}_{cd}} \\
    &= \bar{K}{}\indices{^{ab|}_m} + \delta^{[a}_m \bar{\bar{K}}{}^{b]}\, ,
\end{align}
which reduces to \eqref{eq:dualityAT} on-shell using \eqref{eq:ttG}.

This action is obtained from \cite{Boulanger:2015mka} using the same reasoning: we start from the action \eqref{eq:actionY} in $D=4$. Here, the decomposition of $Y\indices{^{ab}_c}$ analogous to \eqref{eq:Ydecomp3D} reads
\begin{equation}
    Y^{ab|}{}_{c} = \tfrac{1}{\sqrt{2}} \,\varepsilon^{abmn} T_{mn|c} 
    + 2\,\delta^{[a}_{c} Z^{b]}\; ,
\end{equation}
with the first term being traceless because of the symmetries of $T_{ab|c}$. The resulting action is invariant under the gauge symmetries
\begin{equation}
    \delta T_{ab|c} = 2 \,\pd_{[a} a_{b]c} + 2 \,\pd_{[a} s_{b]c} - 2 \,\pd_c a_{ab}\, , \quad \delta Z_a = \pd_a \xi - 2\sqrt{2}\, \varepsilon_{abcd} \pd^b a^{cd}\, .
\end{equation}
Because of its $\xi$ gauge symmetry, $Z_a$ only appears in the action through its field strength $F_{ab}[Z]$. It can therefore be dualised in the usual way into another vector field $B_{a}$, which however still transforms with the gauge parameters of $T_{ab|c}$,
\begin{equation}
    \delta B_a = \pd_a \lambda -\sqrt{2}\; \pd^b\left( 3\, a_{ab} + s_{ab}\right) \, .
\end{equation}
The last step is to disentangle the gauge transformations by the field redefinition
\begin{equation}
    A_a = B_a - \sqrt{2}\, T\indices{_{ab}^b}\, .
\end{equation}
This gives the Lagrangian \eqref{eq:lagAT4D}, with independent gauge transformations \eqref{gaugeA4D} and \eqref{gaugeT4D}.

\subsection{Double-dual graviton in \texorpdfstring{$D=5$}{D=5}}

We finish this section with the example of the double dual graviton in five spacetime dimensions. It is a $[2,2]$ mixed symmetry field `window' $C_{ab|cd}$, i.e.
\begin{equation}
    C_{ab|cd} = -C_{ba|cd} = -C_{ab|dc} \, , \quad C_{[ab|c]d} = 0
\end{equation}
(algebraic symmetries of the Riemann tensor). 
Its gauge transformation law is
\begin{equation}\label{eq:deltaC}
\delta C^{ab|}{}_{cd} \,=\,2\,\partial^{[a} m_{cd|}{}^{b]}
 + 2\,\partial_{[c} m^{ab|}{}_{d]}\, ,
\end{equation}
where $m_{ab|c}$ is a $[2,1]$ `hook' mixed symmetry tensor, obeying
\begin{equation}\label{mcond}
m_{ab|c} = - m_{ba|c}\, \quad m_{[ab|c]} = 0\, .
\end{equation}
The equation of motion of the double-dual graviton is the double-trace condition
\begin{equation}
    \bar{\bar{K}}_{mn}[C] = 0
\end{equation}
on its gauge-invariant curvature tensor
\begin{equation}
    K^{abc|}{}_{mnp}[C] = 9 \,\partial^{[a} 
    \partial_{[m} C^{bc]}{}_{np]}\, .
\end{equation}
This follows from the double-duality relation with the linearized Riemann tensor of the usual graviton $h_{mn}$, which can be written as
\begin{align}
    \cR^{ab}{}_{cd}[h] &= -\tfrac{\kappa}{36} \varepsilon^{abm_1m_2m_3} \varepsilon_{cdn_1n_2n_3} K\indices{_{m_1m_2m_3|}^{n_1n_2n_3}}[C] \\
    &= \kappa \left( \bar{K}\indices{^{ab|}_{cd}} - 2\, \delta^{[a}_{[c} \bbar{K}{}^{b]}_{d]} + \tfrac{1}{3} \delta^{ab}_{cd} \bbbar{K} \right)
\end{align}
up to a conventional factor $\kappa$. From this equation, it then follows that $\bar{\cR}_{mn}[h] = 0$ is equivalent to $\bbar{K}_{mn}[C] = 0$. Given the result of the previous examples, one can guess the structure of an action giving these equations: the difference of the conventional Lagrangian for the $C$-field and the Fierz-Pauli action, accompanied by a gauge-invariant cross-term,
\begin{equation}\label{eq:doubleduallag}
S[C_{ab|cd}, h_{ab}] = \int\!d^5\!x \left[ \alpha \left(\cL_{[2,2]} - \cL_\text{FP} \right)+ \beta \cL_\text{cross} \right]
\end{equation}
with $\alpha = \pm 1$ and $\beta\neq 0$.
The Lagrangian for the $[2,2]$ field reads \cite{Burdik:2000kj}
\begin{equation}\label{eq:L22}
    {\cal L}_{[2,2]} = ~\tfrac{1}{16}\,\partial_{c}C_{ab|}{}^{ij}\,
    \partial^{k}C_{de|}{}^{lm}\,\varepsilon^{abcde}\,\varepsilon_{ijklm}
\end{equation}
and was first written down in this explicit form in \cite{Bekaert:2004dz} 
by generalising an observation 
done in a topological case in \cite{Curtright:1980yk}.
The cross-term (with two derivatives) is uniquely determined by the gauge transformations \eqref{eq:deltaC} and $\delta h_{ab} = 2 \pd_{(a} \epsilon_{b)}$ up to integration by parts, and can be written as
\begin{equation}
    \cL_\text{cross} = h_{ab} \left( \bbar{K}{}^{ab} - \tfrac{1}{3} \eta^{ab} \bbbar{K} \right) \, ,
\end{equation}
where the quantity in parentheses is divergenceless because of the Bianchi identity satisfied by $K\indices{^{abc}_{mnp}}[C]$. Indeed, this reproduces the required equations of motion and duality relations
\begin{equation}
    \bbar{K}_{ab}[C] = 0 = \bar{\cR}_{ab}[h]\,, \quad  \cR\indices{^{ab}_{cd}}[h] = \kappa\, \bar{K}{}\indices{^{ab|}_{cd}}[C]\, ,
\end{equation}
with $\kappa = \alpha/\beta$, as long as $\beta$ does not take the values $0$ or $\pm 3/(4 \sqrt{2})$ (the sign $\alpha$ could in principle be determined by a Hamiltonian analysis of \eqref{eq:doubleduallag}).

We show now that it is indeed this structure that comes out of the off-shell duality procedure of \cite{Boulanger:2012df} for the double-dual graviton, with the result $\alpha = -1$ and $\beta = -\sqrt{3}/2$. It provides 
a dual action in terms of a gauge potential $D\indices{^{abc|}_{de}}$ which is antisymmetric in both groups of indices, but satisfying no other condition. The action reads
\begin{align}
    S[D\indices{^{abc|}_{de}}] = \frac{1}{4} \int \! d^5\!x\, \Big( &-\pd_a D^{abc|de} \,\pd^f D_{fbc|de} + \pd_a D^{abc|de} \,\pd^f D_{fde|bc} - 2 \pd_a D^{abc|de} \,\pd^f D_{fda|ec} \nonumber\\
    &+3\, \pd_a D\indices{^{ab|}_c} \,\pd^d D\indices{_{db|}^c} - \pd_a D\indices{^{ab|}_c} \,\pd_d D\indices{^{dc|}_b} - \tfrac{1}{3} \pd_a D^a \,\pd_b D^b
    \Big)\, ,\label{eq:actionD}
\end{align}
where the traces are defined as $D\indices{^{ab|}_c} = D\indices{^{abd|}_{cd}}$, $D^a = D\indices{^{abc|}_{bc}}$. Let us review briefly how this action arises by dualising the Curtright action \cite{Boulanger:2012df} (which itself arises from dualising the Fierz-Pauli action in $D=5$ \cite{Boulanger:2003vs}, so this is indeed the case relevant for the double-dual graviton).
The first step is to rewrite the Curtright action (see \eqref{eq:curtrightlag}), up to integration by parts, as
\begin{align}
    S_C[T_{ab|c}] &= \int d^5\!x \,\Big( \tfrac{1}{2} H_{ab|cd}[T] H^{ab|cd}[T] + H_{ab|cd}[T] H^{ac|bd}[T] - 3\,H\indices{_{ac|b}^c}[T] H\indices{^{ad|b}_d}[T] \label{eq:curtH}\\
    &\qquad\qquad - H\indices{_{ac|b}^c}[T] H\indices{^{bd|a}_d}[T] + H\indices{_{ab|}^{ab}}[T] H\indices{_{cd|}^{cd}}[T] \Big) \nonumber \\
    H\indices{^{ab|}_{cd}}[T]&:= 2\, \pd_{[c} T\indices{^{ab|}_{d]}}\, .
\end{align}
Here, the field $T_{ab|c}$ can be assumed to also contain a totally antisymmetric part which is pure gauge; this action is then invariant under $\delta T_{ab|c} = \pd_{[a} \xi_{b]c} + \tfrac{1}{2}\Lambda_{abc}$, where $\Lambda$ is totally antisymmetric and $\xi$ has no particular symmetry. Then, the parent action is
\begin{equation}
    S[H\indices{^{ab|}_{cd}}, D\indices{^{abc|}_{de}}] = \int d^5\!x \,\left( \cL(H) - H\indices{^{ab|}_{cd}} \pd_e D\indices{^{cde|}_{ab}} \right)\, ,
\end{equation}
where $H\indices{^{ab|}_{cd}}$ is here an independent field with two groups of antisymmetric indices, and $\cL_H$ can be read from \eqref{eq:curtH} but without imposing $H = H[T]$. As usual, one can then choose which field to eliminate from this parent action. On the one hand, the equation of motion for $D$ gives the relation $H = H[T]$ using the Poincaré lemma: this then gives back the original Curtright action \eqref{eq:curtH}. On the other hand, the equations of motion for $H$ can be solved algebraically as
\begin{equation}
    H\indices{_{ab|}^{cd}} = \tfrac{1}{2} \left( \pd_e D\indices{^{cde|}_{ab}} - \pd^e D\indices{_{abe|}^{cd}} \right) + \pd_e D\indices{^{e[c}_{[a|}^{d]}_{b]}} - \tfrac{3}{2} \pd_e D\indices{^{e[c}_{[a|}} \delta^{d]}_{b]} + \tfrac{1}{2} \pd^e D\indices{_{e[a|}^{[c}} \delta^{d]}_{b]} + \tfrac{1}{6} \delta_{ab}^{cd} \pd_e D^e
\end{equation}
(we correct a typo in the last coefficient with respect to \cite{Boulanger:2012df}). Plugging this back in the parent action leads to the action \eqref{eq:actionD}. That action is invariant under
\begin{align}
    \delta D\indices{^{abc|}_{de}} &= \pd_f \psi\indices{^{abcf|}_{de}} - 3\, \delta^{[a}_{[d} \Lambda\indices{^{bc]}_{e]}} \\
    &\quad -6 \left( 3\,\delta^{[a}_{[d}\pd_{e]} \xi^{bc]} - \delta^{[a}_{[d}\pd^{b} \xi\indices{^{c]}_{e]}} - 3\,\delta^{[a}_{[d}\pd^{b} \xi\indices{_{e]}^{c]}} + 2\, \delta^{[a}_{[d} \delta^{b}_{e]} \pd_f \xi^{c]f} - 2\,\delta^{[a}_{[d} \delta^{b}_{e]} \pd^{c]} \xi\indices{_f^f}\right)\, ,\nonumber
\end{align}
where $\Lambda$ and $\xi$ come from the invariances of the original action \eqref{eq:curtH}, and the new parameter $\psi$ (antisymmetric in both groups of indices) arises because $D\indices{^{abc|}_{de}}$ only appears through its divergence.

As in the previous sections, we now decompose $D\indices{^{abc|}_{de}}$ into traceless and traceful parts,
\begin{equation}\label{eq:decompD}
    D\indices{^{abc|}_{de}} = - \frac{1}{2} \,\varepsilon^{abcij} \,C_{de|ij} + 6\, \delta^{[a}_{[d} Z\indices{^{bc]|}_{e]}}\, .
\end{equation}
Here, $C_{ab|cd}$ is the $[2,2]$ double dual graviton field, 
and the gauge field $Z\indices{^{ab|}_c}$ obeys $Z\indices{^{ab|}_c}=-Z\indices{^{ba|}_c}\,$ but no other condition\footnote{In the notation of \cite{Boulanger:2012df}, $C_{ab|cd}$ is the Hodge dual of $X\indices{^{ijk}_{cd}}$ on its first three indices, and $Z\indices{^{ab|}_c}$ is the traceful combination $Z\indices{^{(1)ab}_c} + \delta^{[a}_c Z^{(2)b]}$}.
Using this decomposition in \eqref{eq:actionD} yields an action
\begin{equation}\label{eq:SCZ}
    S[C_{ab|cd},Z\indices{^{ab|}_c}] = \int \!d^5\!x\, \left[ {\cal L}(\partial C) + {\cal L}(\partial Z)  + {\cal L}^{\rm cross}(\pd C, \pd Z) \right]
\end{equation}
that contains all the components of $C_{ab|cd}$. Without loss of generality, it is also useful to write the gauge parameter $\psi\indices{^{abcd|}_{ef}}$ as
\begin{equation}
    \psi\indices{^{abcd|}_{ef}} = \varepsilon^{abcdg} \left( -2\, m_{ef|g} + \varepsilon_{efgij}\, a^{ij} \right)
\end{equation}
in terms of a $[2,1]$ gauge parameter $m_{ab|c}$ obeying
\eqref{mcond}, and an antisymmetric gauge parameter $a^{ij}$. The latter can be absorbed by a redefinition of $\Lambda^{abc}$ and (the antisymmetric part of) $\xi^{ab}$, so that $\psi\indices{^{abcd}_{ef}}$ is fully replaced by $m_{ab|c}$. The resulting gauge transformations can then be written as
\begin{align}
    \delta C_{ab|cd} \,=\, & \,2\,\partial_{[a|}m_{cd|b]}+
    2\,\partial_{[c|}m_{ab|d]}\;,\\
    \delta Z^{ab|}{}_{c} \,=\, &\,\Lambda^{ab}{}_c 
                                 + \partial^{[a}\xi^{b]}{}_c
                                 - \tfrac{1}{2}\, \delta^{[a}_c\,\partial_{d}\xi^{b]d} +\tfrac{1}{2}\varepsilon\indices{^{ab}_{pqr}} \pd^p m\indices{^{qr}_c}\, .
\end{align}
In particular, one finds the expected gauge transformations \eqref{eq:deltaC} for the double dual graviton field. However, since the gauge parameter $m_{ab|c}$ also appears in the transformation of $Z\indices{^{ab}_c}$, the term $\cL(\pd C)$ appearing in \eqref{eq:SCZ} is \emph{not} the usual Lagrangian \eqref{eq:L22} for a $[2,2]$ mixed symmetry field. To make contact with \eqref{eq:L22}, one must (following the logic of the previous sections) first dualise the field $Z\indices{^{ab|}_c}$. To do this, we use the field variable
\begin{align}
    Y^{ab|}{}_c := Z^{ab|}{}_c + \delta^{[a}_c Z^{b]}\;
\end{align}
that transforms as
\begin{align}
    \delta Y^{ab|}{}_c =& \,\Lambda^{ab}{}_c 
                                 + \partial^{[a}\left( \xi^{b]}{}_c -\tfrac{1}{2} \delta^{b]}_c \xi\indices{^d_d} \right)+\tfrac{1}{2}\,\varepsilon\indices{^{ab}_{pqr}} \pd^p m\indices{^{qr|}_c}\;.
\end{align}
Indeed, the invariance under $\xi_{ab}$ gauge transformations then implies that the Lagrangian depends on $Y$ only through the quantity 
$F^{abc|}{}_{d}(Y):=3\,\partial^{[a}Y^{bc]|}{}_d\,$:
\begin{align}
    {\cal L}(\partial C,\partial Y) \,=\,&\,  -\tfrac{1}{6}\, F^{abc|d}(Y)\,F_{abc|d}(Y)
    -\tfrac{1}{4}\, F^{abc|d}(Y)\,F_{dab|c}(Y) 
    + \tfrac{3}{4}\, F^{abc|}{}_{c}(Y)\,F_{abd|}{}^{d}(Y)
    \nonumber \\
    & +{\cal L}(\partial C)
    + \tfrac{1}{2}\,\varepsilon_{mnpqr}\,
    \partial^{p}C^{qr|}{}_{ab}\,
    \left[ F^{mna|b}(Y)  - \tfrac{1}{2}\, F^{abm|n}(Y)\right]\;.
\label{dualCZ}
\end{align}
Note that, using the $\Lambda\indices{^{ab}_c}$ gauge transformation, the totally antisymmetric part of $Y\indices{^{ab|}_c}$ could be gauged away and $Y\indices{^{ab|}_c}$ taken to be an irreducible $[2,1]$ (Curtright) field. The first line of \eqref{dualCZ} would then reduce to the Curtright Lagrangian \eqref{eq:curtrightlag}.

Now, by the standard procedure one can trade the field $Y^{ab|}{}_c\,$, 
--- i.e., $Z^{ab|}{}_c\,$ --- for the Fierz-Pauli symmetric 
rank-two potential $h_{mn}$ by introducing the parent Lagrangian
\begin{align}
    {\cal L}(\partial C,\partial f,F) \,=\, &\,  -\tfrac{1}{6}\, F^{abc|d}\,F_{abc|d}
    -\tfrac{1}{4}\, F^{abc|d}\,F_{dab|c} 
    + \tfrac{3}{4}\, F^{abc|}{}_{c}\,F_{abd|}{}^{d} +{\cal L}(\partial C)
    \nonumber \\
    & + \tfrac{1}{2}\,\varepsilon_{mnpqr}\,
    \partial^{p}C^{qr|}{}_{ab}\,
    \left( F^{mna|b}  - \tfrac{1}{2}\, F^{abm|n}\right)
    + \varepsilon_{abcpq}\,\partial^{p}f^{q}{}_{d}\,
    F^{abc|d}\;,
    \label{parentCfF}
\end{align}
where $F^{abc|d}$ is viewed as an independent field and 
where $f_{mn}$ contains also an antisymmetric component. 
The equations of motion for $f^{mn}$ imply that $F=F(Y)\,$, which 
correctly reproduces \eqref{dualCZ}. On the other hand, the 
field $F$ is an auxiliary field inside the action 
$S[C,f,F]=\int d^5x\,\cL(\partial C,\partial f,F)\,$. 
Its field equations can be solved algebraically to yield
\begin{align}
    F_{abc|d} =& ~\varepsilon_{abcmn}\,
    (\partial^{m}f^{n}{}_{d}+\partial^{m}f_{d}{}^{n} 
    -\partial_{d}f^{mn})
    +2\,\varepsilon_{abcdm}
    (\partial^{m}f^{n}{}_{n}-\partial^{n}f^{m}{}_{n})
    \nonumber \\
    & + \tfrac{1}{2}\,\varepsilon_{abcmn}\,
    (\partial_p C^{mn|p}{}_{d}
    -2\,\partial^{m}C^{n}{}_{d})
    + \tfrac{1}{2}\,\varepsilon_{abcdm}\,
    (2\,\partial_{n}C^{mn}-\partial^{m}C)\;.
    \label{solutionforF}
\end{align}
Upon substituting this expression for $F_{\mu\nu\rho|\sigma}$ inside the action 
$S[C,f,F]\,$, we find the action
\begin{equation}
    S[C,f]= \int d^5x\,\left[ {\cal L}_{[1,1]}(\partial f) - {\cal L}_{[2,2]}(\partial C)
    -\tfrac{3}{2}\,f^{ab}\,G_{ac|}{}_b{}^c[C]\right]\;,
\end{equation}
where
\begin{align}
    {\cal L}_{[1,1]}(\partial f) &= -\tfrac{3}{2}\,\left(
    \Omega^{ab|c}\Omega_{ab|c}+2\,\Omega^{ab|c}\Omega_{ac|b}
    -4\,\Omega_{ab|}{}^{b}\Omega^{ac|}{}_{c}\right)\;,\quad \Omega_{ab|c}:=2\,\partial_{[a}f_{b]c}\;,
    \\
    {\cal L}_{[2,2]}(\partial C) &= ~\tfrac{1}{16}\,\partial_{c}C_{ab|}{}^{ij}\,
    \partial^{k}C_{de|}{}^{lm}\,\varepsilon^{abcde}\,\varepsilon_{ijklm}\,\;,
    \\
    G_{ab|}{}^{ij}[C] &= \varepsilon_{abcde}\,\varepsilon^{ijklm}\,\partial^e\partial_m\,
    C^{cd|}{}_{kl} = \tfrac{1}{9} \varepsilon_{abcde}\,\varepsilon^{ijklm} K\indices{_{klm|}^{cde}}[C]\;.
\end{align}
The antisymmetric part of $f_{ab}$ does not appear in this action, and $\cL_{[1,1]} (\pd f)$ is the usual Fierz-Pauli action up to a total derivative (and the rescaling $f_{ab} = h_{ab}/2\sqrt{3}$ needed to reproduce the normalisation of \eqref{eq:fierzpauli}). The Lagrangian $\cL_{[2,2]}$ is indeed \eqref{eq:L22}. Finally, due to the trace identity
\begin{equation}
    G_{ac|}{}_b{}^c = 2 \left( \bbar{K}_{ab} - \tfrac{1}{3} \eta_{ab} \bbbar{K}\right)\, ,
\end{equation}
we have indeed recovered the structure \eqref{eq:doubleduallag} with the special value $\beta = -\sqrt{3}/2$.

\section{Three-columns in three dimensions}
\label{sec:threecolumns}

\subsection{Topological case: Spin 2 -- Spin 3}\label{sec:3Dtop}

In this section, we consider the dualisation of a spin two field $h_{ab}$ in three space-time dimensions on an empty column, which leads to a spin three gauge field $\phi_{abc}$. The off-shell dualisation procedure, starting from the topological linearised gravity action for $h_{ab}$, yields a novel topological action in $D=3$ involving both fields. (This is to be contrasted with the dualisation of $h_{ab}$ on a non-empty column, which gives the topological Lagrangian $\cL = 0$ \cite{Boulanger:2003vs}.)

\paragraph{Relation between curvatures.}

Seen as a degenerate $[1,1,0]$ field, the natural curvature of $h_{ab}$ is the five-index object
\begin{equation}
    K_{ab|cd|e}[h] = 4\, \pd_e \pd_{[a} \pd_{[c} h_{d]b]} = \pd_e \cR_{ab\,cd}[h] \sim \ydiagram{3,2}\, ,
\end{equation}
where $R$ is the linearized Riemann tensor of $h$. Taking the Hodge dual on its last index, we obtain the duality relation with a spin three field $\phi_{abc}$,
\begin{equation}
K_{ab|cd|ef}[\phi] := 8 \,\pd_{[a} \pd_{[c} \pd_{[e} \phi_{f]d]b]} = K_{ab|cd|p}[h] \,\varepsilon^p{}_{ef}\sim \ydiagram{3,3}\, ,
\end{equation}
where $K[\phi]$ is the natural higher-spin curvature of $\phi_{abc}$. This duality relation alone implies the equations of motion for the fields. In the case of three dimensions, the fields are topological: the equations of motion are equivalent to the vanishing of the curvatures themselves, which then implies that the fields are pure gauge and carry no local degrees of freedom.

\paragraph{Dual action and gauge invariances.}

We start from the Fierz-Pauli action, under the form
\begin{align}
    S_\text{FP}[h_{ab}] = \int \!d^3x
    \left[-\tfrac{1}{2}\,\partial_a h_{bc}\,\partial^a h^{bc}\,
    +\tfrac{1}{2}\,\partial_a h\,\partial^a h\,
    - \partial_a h^{ab}\,\partial_b h + \partial_a h^{ab}\,\partial^c h_{cb} \right]\;.
\end{align}
Following \cite{Boulanger:2012df}, we define the parent action
\begin{align}
    S[G_{a|bc},D_{ab|}{}^{cd}] &= \int \!d^3x
    \left[-\tfrac{1}{2}\,G_{a|bc}\,G^{a|bc}+\tfrac{1}{2}\,G_{a|c}{}^c \,G^{a|b}{}_b
    - G_{a|}{}^{ab}\,G_{b|c}{}^c + G_{a|}{}^{ab}\,G^{c|}{}_{cb}
    + G\indices{^{d|}_{bc}}\,\partial^{a}D\indices{_{ad|}^{bc}}\right]
\end{align}
that contains the fields $G_{a|bc}$ and $D_{ab|}{}^{cd}$, with index symmetries
\begin{align}
G_{a|bc} &= +G_{a|cb} \\
D_{ab|}{}^{cd} &= -D_{ba|}{}^{cd}=+D_{ab|}{}^{dc}\, .
\end{align}
This action is invariant under the gauge transformations
\begin{align}
    \delta G\indices{^{a|}_{bc}} &= 2 \,\pd^a \pd_{(b} \epsilon_{c)} \;,\\
    \delta D\indices{_{ab|}^{cd}} &= \varepsilon_{abp} \,\pd^p \upsilon^{cd} 
    + 2\,\eta^{cd} \pd_{[a} \epsilon_{b]}+ 4\,\delta_{[a}^{(c} \pd_{b]}
    \epsilon^{d)}\;,
\end{align}
where $\upsilon^{cd}$ is symmetric and $\epsilon_a$ is a vector parameter.

The equation of motion of $D$ gives $\pd^{[a} G\indices{^{b]|}_{cd}} = 0$, which can be solved as $G\indices{^{a|}_{bc}} = \pd^a h_{bc}$ for some symmetric tensor $h_{ab}$. Plugging this solution back into the parent action then brings us back to the original Fierz-Pauli action. On the other hand, the equation of motion for $G_{a|bc}$, which reads
\begin{equation}
    -G_{a|bc} + \eta_{bc}G\indices{_{a|}^d_d} - \eta_{a(b}G\indices{_{c)|}^d_d} - \eta_{bc} G\indices{^{a|}_{bd}}+ 2 \eta_{a(b}G\indices{^{d|}_{c)d}} + \pd^d D_{da|bc} = 0\, ,
\end{equation}
can be solved algebraically for $G_{abc}$ as
\begin{equation}
    G_{a|bc} = \pd^d D_{da|bc}-\eta_{bc} \pd^d D\indices{_{da|e}^e} - \eta_{a(b|} \pd_d D\indices{^{de|}_{c)e}}\, .
\end{equation}
Plugging this back into the parent action gives the dual action
\begin{equation}\label{eq:topactionD}
S[D_{ab|cd}] =\frac{1}{2} \int \!d^3x\, \left( 
\partial^a D_{ab|cd} \,\partial_e D^{eb|cd} 
- \partial^a D\indices{_{ab|c}^b} \partial_d D\indices{^{de|c}_e} 
- \partial^a D\indices{_{ab|c}^c} \partial_e D\indices{^{eb|d}_d} \right)   
\end{equation}

Since we are in three dimensions, the $\varepsilon^{abc}$ tensor 
can be used to trade a pair of antisymmetric indices for a single index; accordingly, we will also use the three-index tensor
\begin{equation}
    \widetilde{D}^{a |ij} := - \frac{1}{2} \varepsilon^{abc} D\indices{_{bc|}^{ij}} \qquad \Leftrightarrow \qquad D\indices{_{ab|}^{ij}} = \varepsilon_{abc} \widetilde{D}^{c|ij}
\end{equation}
instead of $D\,$. One also introduces 
\begin{align}
D_{ab|}{}^{ij} = X_{ab|}{}^{ij} + 4\, \delta^{(i}{}_{[a}\,Z_{b]}{}^{j)}  \;,
    \quad X_{ab|}{}^{ib}\equiv 0 \equiv Z_{a}{}^a\;,
\end{align}
with inverse formulas
\begin{align}
    X\indices{_{ab|}^{ij}} = D\indices{_{ab|}^{ij}} + \frac{4}{3}\delta^{(i}_{[a}\,D\indices{_{b]k|}^{j)k}}\; , \quad Z\indices{_a^i} = - \tfrac{1}{3} D\indices{_{ak|}^{ik}}\; .
\end{align}
In terms of the symmetric rank-3 field 
\begin{align}
    \widetilde{\varphi}^{abc}:= -\tfrac{1}{2}\,\varepsilon^{aij}\,X_{ij|}{}^{bc}
    \qquad \Leftrightarrow \qquad 
    X_{ab|}{}^{ij} = \varepsilon_{abc}\,\widetilde{\varphi}^{ijc}\;,
\end{align}
one has
\begin{align}
   \widetilde{D}^{a |ij} = \widetilde{\varphi}^{aij} 
   + 2\, \varepsilon^{ab(i}\,Z_b{}^{j)}
   \qquad \Leftrightarrow \qquad 
   \widetilde{\varphi}^{aij} = \widetilde{D}^{(a |ij)}\;,\quad 
   Z_m{}^j = \tfrac{1}{3}\,\varepsilon_{abm}\,\widetilde{D}^{a| bj}\;.
\end{align}
The gauge transformation laws then read
\begin{align}
    \delta\widetilde{D}_{a|bc} &= \pd_a \upsilon_{bc} 
    + 2\, \varepsilon_{ap(b} \pd^p \epsilon_{c)} 
    - \eta_{bc} \,\varepsilon_{apq} \pd^p \epsilon^q\;,\\
    \delta Z_a{}^i &= \tfrac{1}{3}\,\left( 
    \varepsilon_{abc}\partial^b\upsilon^{ci}
    +2\,\partial_a\epsilon^i +\partial^i\epsilon_a 
    -\delta^i_a\,\partial_b \epsilon^b
    \right)\;.
\end{align}

\paragraph{Change of variables.}

We define a symmetric field $\phi_{abc}\,$, a traceless symmetric field 
$f_{ab}$ and a vector $A_a$ as follows:
\begin{align}
    \phi_{abc} &:= \widetilde{D}_{(a|bc)} - 
    \eta_{(ab} \widetilde{D}\indices{_{c)|i}^i} 
    =
    \widetilde{\varphi}_{abc} - \eta_{(ab}\,\widetilde{\varphi}_{c)} 
    - 2\,\eta_{(ab}\,\varepsilon_{c)ij}\, Z^{ij}\;,
    \\
    f_{ab} &:= \varepsilon_{pq(a} \widetilde{D}\indices{^{p|q}_{b)}}
    \quad \Leftrightarrow \quad 
    f_{ab} = 3\, Z_{(ab)}\;,
    \\
    A_a &:= - \widetilde{D}\indices{_{a|b}^b} 
    \quad \Leftrightarrow \quad 
    A^a = - \widetilde{\varphi}^a - 2\,\varepsilon^{abc}\,Z_{bc} = 
    \tfrac{3}{2}\,\phi^a + 3\,\varepsilon^{abc}\,Z_{bc}  \;.
\end{align}
The inverse formula, expressing $\widetilde{D}$ 
as a function of $(\phi_{abc}, f_{ab}, A_a)$, is 
\begin{equation}
    \widetilde{D}_{a|bc} = \phi_{abc} 
    - \tfrac{2}{3}\,\varepsilon_{a(b}{}^m\,f_{c)m} 
    + \tfrac{1}{2}\,\eta_{a(b}\,\phi_{c)} 
    - \tfrac{1}{2}\,\eta_{bc}\,\phi_{a}
    - \eta_{a(b}\,A_{c)}\;.
\end{equation}
In terms of these fields, we get an action 
invariant under
\begin{align}
    \delta \phi_{abc} &= \pd_{(a} \xi_{bc)}\;, \qquad
    \label{GT1}
    \\
    \delta f_{ab} & = 3\, \pd_{(a}\epsilon_{b)} 
    - \eta_{ab} \pd_c \epsilon^c 
    + \varepsilon_{mn(a} \partial^m \xi\indices{^n_{b)}} \;,
    \\
    \delta A_a &= \tfrac{1}{2}\,\pd_a {\xi} 
    + \varepsilon_{apq} \pd^p \epsilon^q \;,\label{Utransfo}
\end{align}
where we redefined the gauge parameter $\upsilon_{ab}$ as
\begin{equation}
    \xi_{ab} := \upsilon_{ab} - \eta_{ab} \,{\upsilon}
    \qquad \Rightarrow \qquad \xi = -2\,{\upsilon}\;.
\end{equation}
In the formulae \eqref{GT1}, 
the transformation of $\phi_{abc}$ is the usual one for a spin three 
massless field, except for the fact that the trace of the gauge parameter 
$\xi_{ab}$ is not identically zero. 
The trace $\xi:=\xi^a{}_a$ appears also in the transformation law 
of the vector field, in \eqref{Utransfo}. 
Upon changing variables from $\phi_{abc}$ to 
\begin{align}
    \varphi_{abc} := \phi_{abc}-\tfrac{2}{3}\,\eta_{(ab}\,A_{c)}\;,
\end{align}
we have that this field transforms according to 
\begin{align}
    \delta \varphi_{abc} = \partial_{(a}\widehat{\xi}_{bc)} 
    - \tfrac{2}{3}\,\varepsilon_{(a}{}^{pq}\,\eta_{bc)} \pd_p \epsilon_q\;,
\end{align}
where $\widehat{\xi}_{ab} := \xi_{ab} - \tfrac{1}{3}\,\eta_{ab}\,\xi\,$
is the traceless part of $\xi_{ab}\,$.
After trading $\phi_{abc}$ for $\varphi_{abc}$, the only field 
that transforms with the trace of $\xi_{ab}$ is the vector $A_a\,$
that must therefore appear in the action only through its field strength
$F_{ab}[A]=2\,\partial_{[a}A_{b]}\,$. We can then dualise the vector 
$A_a$ into a scalar $\sigma\,$. 
For this we add the term $\varepsilon_{abc}F^{ab}\partial^c\sigma\,$ 
to the Lagrangian where $F^{ab}[A]$ is replaced by the independent 
antisymmetric tensor field $F^{ab}\,$. Extremising with respect to the 
auxiliary field $F^{ab}$ enables one to eliminate it in terms of the 
other fields, giving an action $S[\varphi_{abc},f_{ab},\sigma]\,$ that 
is invariant under 
\begin{align}
    \delta \varphi_{abc} &= \partial_{(a}\widehat{\xi}_{bc)} 
    - \tfrac{2}{3}\,\varepsilon_{(a}{}^{pq}\,\eta_{bc)} \pd_p \epsilon_q\;,
    \\
    \delta f_{ab} &= 3\, \pd_{(a}\epsilon_{b)} 
    - \eta_{ab} \pd_c \epsilon^c 
    + \varepsilon_{mn(a} \partial^m \xi\indices{^n_{b)}} \;,
    \\
    \delta \sigma &= -\tfrac{2}{3}\,\partial_a\epsilon^a\;. 
\end{align}
One can then combine the two fields $f_{ab}$ and $\sigma$ into a 
traceful tensor $h_{ab} := \tfrac{2}{3}\,f_{ab}- \eta_{ab}\,\sigma\,$, giving the following action:
\begin{align}
    S[\varphi_{abc},h_{ab}] = \frac{1}{2}\,\int d^3x \Big[&
    -\partial_a\varphi_{bcd}\,\partial^a\varphi^{bcd}+
    \partial^a\varphi^b\,\partial^c\varphi_{abc}
    +\partial_a\varphi^{abc}\,\partial^d\varphi_{bcd}
    \nonumber \\
    & -\tfrac{1}{7}\,\partial_a \varphi_b\,\partial^a \varphi^b
    -\tfrac{31}{28}\,\partial_a \varphi^a\,\partial^b \varphi_b
    \nonumber \\
    & + \tfrac{1}{2}\,\partial_a h_{bc}\, \partial^a h^{bc}
    + \tfrac{1}{14}\,\partial_a h\, \partial^a h
    - \tfrac{3}{7}\,\partial^a h_{ab}\, \partial_c h^{bc}
    - \tfrac{1}{7}\,\partial^a h\, \partial_c h_a{}^c
    \nonumber \\
    & + \tfrac{10}{7}\,\varepsilon_{apq}\,\partial^b h_{b}{}^a\,
    \partial^p\varphi^q 
    - 2\,\varepsilon_{apq}\,\partial^b h^{ac}\,
    \partial^p\varphi^q{}_{bc} 
    \Big]
\end{align}
that is invariant under 
\begin{align}
    \delta \varphi_{abc} &= 3\,\partial_{(a}\widehat{\xi}_{bc)} 
    - \tfrac{2}{3}\,\varepsilon_{(a}{}^{pq}\,\eta_{bc)} \pd_p \epsilon_q\;,
    \\
    \delta h_{ab} &= 2\, \pd_{(a}\epsilon_{b)} 
    + {2}\,\varepsilon_{pq(a} \partial^p \widehat{\xi}^q{}_{b)} \;.
\end{align}
A general result from \cite{Grigoriev:2020lzu} states that 
the above action should be expressible in a Chern-Simons form, 
although this is not straightforward from the entangled form 
of the gauge transformations.
We hope to report about this point in the near future \cite{inprogress}.

\paragraph{Analysis of the degrees of freedom.}
In this subsection, we prove that the theory described here is topological. Of course, this is a consequence of the construction since it is equivalent off-shell to the Fierz-Pauli theory. However, it is not an obvious fact when looking only at the final form of the action and gauge transformations; for completeness, in what follows we present an alternative proof of this fact.
In order to understand the nature of the physical degrees of freedom of a free theory, the strategy we will follow is to list all the gauge-invariant quantities that do not vanish on shell and identify the differential equations they satisfy. 
If it happens that all the gauge-invariant quantities vanish on shell, the theory 
is topological: there is not propagating degree of freedom. 

The classification of gauge-invariant quantities boils down to a purely 
algebraic problem in the jet space of the fields, the gauge parameters 
and all their derivatives. From the very structure of the gauge transformations 
at hand, \eqref{GT1}--\eqref{Utransfo}, it appears that one must decompose
into $so(3)$-irreps the $n$th derivatives of the gauge fields and compare the 
resulting set of irreps with the $so(3)$-irreducible decomposition of the 
space of $(n+1)$th derivatives of the gauge parametres.

If some irreps appear in the first list that have no equivalent in the second 
list, these irreps indicate the existence of the gauge-invariant combinations
built out of the $n$th derivatives of the gauge fields. 
If some $so(3)$-irreps appear on the second list that have no equivalent 
counterpart in the first one, these $so(3)$-irreps indicate the existence 
of linear combinations of the $(n+1)$th derivatives of the gauge parameters
that cannot be written as the gauge variation of a linear combination of 
the $n$th derivatives of the gauge fields. 

Clearly, we already know that, at second order in the derivatives of the 
fields, some gauge-invariant quantity will appear: these are the left-hand-side
of the Euler-Lagrange field equations. However, by definition these gauge 
invariant linear combinations of the second derivatives of the fields
vanish on shell. 
What we will show is that all the gauge-invariant quantities
in the theory at hand are of this type. They all vanish on shell, 
so that the theory is topological indeed. 
In the proof that follows, we will denote by $[s]$ the 
spin-$s$ irrep of $so(3)\,$ (of dimension $2s+1\,$) and will systematically 
use the well-known rule 
$[j]\otimes[j']\sim \bigoplus_{s=|j-j'|}^{j+j'} [s]\,$ for 
the addition of angular momenta in 3D. 

\begin{itemize}
    \item There are no linear combinations of the undifferentiated 
    gauge parameters $\Xi = \{ \xi_{ab},\epsilon_a \}$
    that can be written as the gauge variation of the fields, since the 
    latter brings one derivative of the gauge parameters.
    The collection ${\cal S}_\Xi^{(0)}:= \{ [2],[1],[0]\}$ represents the 
    gauge parameters, where the singlet $[0]$ accounts for the trace 
    $\xi=\eta^{ab}\,\xi_{ab}\,$ ;  
    \item At zeroth order in the derivatives of the gauge fields 
    $\Phi = \{\phi_{abc},f_{ab},A_a\} \sim \{[3]\oplus[1],[2],[1]\}$ we 
    have the collection of $so(3)$-irreps 
    ${\cal S}_\Phi^{(0)} := \{ [3], [2],$  $2\times [1] \}\,$. 
    As for the decomposition of the first derivative 
    $\partial \,\Xi$ of 
    the gauge parameters $\Xi = \{ \xi_{ab},\epsilon_a\}$
    $\sim \{ [2]\oplus [0],[1]\}\,$, we find that the collection 
    generated by the tensor product 
    $\partial \,\Xi\sim [1]\otimes\big([2]\oplus[0]\oplus[1]\big)\,$, 
    is given by ${\cal S}_\Xi^{(1)} := \{[3], 2\times[2], 3\times[1], [0]\}\,$. 
    Comparing ${\cal S}_\Phi^{(0)}$ with ${\cal S}_\Xi^{(1)}$ 
    we see that there are three 
    linearly independent combinations of first derivatives of the gauge 
    parameters that cannot be written as the gauge variation of linear 
    combinations of the gauge fields. These three linear combinations 
    transforms in the $so(3)$-irreps in the collection $\{[2],[1],[0]\}\,$.
    Although they are not needed for our proof, they can be taken to be 
    $\{ 3\, \pd_{(a}\epsilon_{b)} 
    - \eta_{ab} \pd_c \epsilon^c 
    - \varepsilon_{mn(a} \partial^m \xi\indices{^n_{b)}}\, ,
    \partial^b \xi_{ab} + \varepsilon_{amn} \partial^m \epsilon^n \,, 
    \partial^a \epsilon_a \}\,$;
    \item Computing the two collections of $so(3)$-irreps generated by 
    $\partial \Phi$ and $\partial^2\,\Xi\,$, 
    respectively, we see that they coincide. This  
    means (i) that there is no gauge-invariant quantity built out of the 
    first derivatives of the gauge fields, and (ii) that there are no quantities 
    built out of the second derivatives of the gauge parameters that 
    cannot be expressed as the gauge variation of some linear combinations 
    of first-order derivatives of the gauge fields ;
    \item Computing the two collections of $so(3)$-irreps generated by 
    $\partial^2 \Phi$ and $\partial^3\,\Xi\,$, 
    respectively, we see that the first one contains extra $so(3)$-irreps
    compared to the second one, indicating the existence of gauge-invariant 
    quantities at second order in the derivatives of the gauge fields. 
    These extra $so(3)$-irreps are given by $\{ [3],[2],2\times [1]\}\,$
    and represent the left-hand sides of the Euler-Lagrange
    field equations ${\cal E}^{\phi}_{abc}:=\frac{\delta {\cal L}}{\delta \phi^{abc}}\,$,
    ${\cal E}^{f}_{ab}:=\frac{\delta {\cal L}}{\delta f^{ab}}\,$ and 
    ${\cal E}^A_{a}:=\frac{\delta {\cal L}}{\delta A^{a}}\,$. 
    These gauge-invariant quantities vanish on shell, by definition, so that 
    they cannot account for propagating degrees of freedom ;
    \item Computing the collection of $so(3)$-irreps generated by the third-order 
    derivatives of the fields, $\partial^3 \Phi\,$, we know for sure that 
    there will be $so(3)$-irreps representing gauge-invariant quantities 
    given by the first-order derivatives 
    $\partial {\cal E}$ of the left-hand sides of the Euler-Lagrange field
    equations, plus, perhaps, extra gauge-invariant quantities that do not 
    vanish on shell. A subtlety that appears at this order is that the 
    left-hand sides of the Euler-Lagrange field equations are not linearly 
    independent, due to the Noether identities. The latter ensures that there 
    are three linear combinations of the field equations that are identically 
    zero, these linear combinations transforming in the $so(3)$-irreps 
    contained in the collection 
    ${\cal S}_\text{Noether}^{(0)} = \{ [2],[1],[0]\}\equiv {\cal S}_\Xi^{(0)}\,$. 
    This is simply the content of Noether's second theorem. 
    Then, if one uses the symbols $\oplus$ and $\ominus$ for the addition 
    and subtraction of collections of $so(3)$-irreps, 
    respectively\footnote{Note that a collection is not the same as a set, 
    in the mathematical sense. The mathematical notion of ``set'' does not 
    account for repetitions of objects.}, direct computation 
    yields $\partial^3\Phi = \partial {\cal E} + \partial^4\Xi\,$, or 
    ${\cal S}_\Phi^{(3)} = {\cal S}_\Phi^{(1)} 
    \ominus {\cal S}_\text{Noether}^{(0)}\oplus {\cal S}_\Xi^{(4)}
    = {\cal S}_\Phi^{(1)} \ominus {\cal S}_{\Xi}^{(0)}\oplus 
    {\cal S}_\Xi^{(4)}\,$,
    where we used that the collection of $so(3)$-irreps appearing in the 
    decomposition of $\partial {\cal E}$ can be obtained by taking 
    ${\cal E}$ to be naively represented by the collection 
    ${\cal S}_\Phi^{(0)}\,$, \emph{provided} one subtracts 
    to the resulting collection the collection of $so(3)$-irreps 
    ${\cal S}_\text{Noether}^{(0)}\equiv {\cal S}_{\Xi}^{(0)}\,$.
    The equality ${\cal S}_\Phi^{(3)} = {\cal S}_\Phi^{(1)} 
    \ominus {\cal S}_{\Xi}^{(0)}\oplus {\cal S}_\Xi^{(4)}\,$ 
    or equivalently ${\cal S}_\Phi^{(3)} \oplus {\cal S}_{\Xi}^{(0)}$
    $={\cal S}_\Phi^{(1)}\oplus {\cal S}_\Xi^{(4)}$
    shows that, at this (third) order in the derivatives of the fields, 
    all the gauge-invariant quantities vanish on shell.
\end{itemize}
In general, one can readily decompose the $so(3)$-irreps corresponding to the $n$th 
derivatives $\partial^n{\cal E}$ ($n>0$) of the left-hand sides of the
Euler-Lagrange field equations $ \cal E$ by taking the latter to be represented 
by the collection ${\cal S}_\Phi^{(0)}$ \emph{provided} one subtracts to the 
resulting collection the collection of $so(3)$-irreps 
${\cal S}_\text{Noether}^{(n-1)}$ obtained by computing the collection of irreps 
generated by the $n-1$th derivatives $\partial^{n-1}\,\Xi\,$.

In order to complete the proof that the theory at hand is topological,  
we only need to show that the relation 
$\partial^{n+2}\Phi = \partial^{n+2}{\cal E} + \partial^{n+3}\Xi\,$ 
is true, or equivalently, that the following relations between collections
is true:
\begin{align}
 {\cal S}_\Phi^{(n+2)} \oplus{\cal S}_{\Xi}^{(n-1)} &= 
 {\cal S}_{\Phi}^{(n)} \oplus  {\cal S}_{\Xi}^{(n+3)}\;
 \nonumber \\
\Leftrightarrow\quad   {\cal S}_\Phi^{(n+2)} \ominus {\cal S}_{\Phi}^{(n)} &= 
  {\cal S}_{\Xi}^{(n+3)} \ominus {\cal S}_{\Xi}^{(n-1)} \;.
 \label{identity}
\end{align}
Indeed, that would imply that all the gauge-invariant quantities that 
the theory admits vanish on shell.
The left-hand side of \eqref{identity} gives 
$[n+2]\otimes\big( [3]\oplus [2] \oplus 2\times [1]\big)$
while the right-hand side of \eqref{identity} produces 
$\big( [n+3]\oplus [n+1] \big)\otimes\big( [2]\oplus [1] \oplus [0]\big)\,$.
A direct decomposition of both sides shows that they are 
identical, which finishes the proof. 

\subsection{Dynamical case: Spin 1 -- Spin 3}

In three dimensions, dualizing a vector field $A_a$ twice on empty columns gives a spin three field $\phi_{abc}$. Equivalently, this field also arises by dualising a scalar three times on empty columns. The difference with the previous section is that this field now carries one degree of freedom.

\paragraph{Equations of motion and duality relations.}

We consider a free Maxwell field $A_a$ in three spacetime dimensions, with the usual equation of motion $\pd^a F_{ab} = 0$ and gauge symmetry $\delta A_a = \pd_a \lambda$. Seen as a $[1,0,0]$ mixed symmetry field with three columns (two of which are empty), its curvature is the four-index tensor
\begin{equation}
    K_{ab|c|d}[A] = 2 \,\pd_d \pd_c \pd_{[a} A_{b]} = \pd_d \pd_c F_{ab}[A] \sim \ydiagram{3,1}\, ,
\end{equation}
or $K[A] = d^3 A = d_2 d_3 F[A]$ in the index-free notation of 
\cite{DuboisViolette:2001jk,Bekaert:2002dt}. It is traceless on-shell. Hodge duality on the last two columns produces the six-index tensor
\begin{equation}
    \left( \star_2 \star_3 K[A] \right)_{ab|cd|ef} := \varepsilon_{cdp} \,\varepsilon_{efq}\, K\indices{_{ab|}^{p|q}}[A] \sim \ydiagram{3,3}\, .
\end{equation}
This tensor satisfies the Bianchi identities implying the existence of a spin-3 field $\phi_{abc}$ such that
\begin{equation}
    \star_2 \star_3 K[A] = K[\phi]\, ,
\end{equation}
where $K[\phi] = d^3 \phi = d_1 d_2 d_3 \phi$ is the usual higher-spin curvature of $\phi$. In components,
\begin{equation}\label{eq:dualityKFnaive}
    K_{ab|cd|ef}[\phi] := 8 \,\pd_{[a} \pd_{[c} \pd_{[e} \phi_{f]d]b]} = \varepsilon_{cdp} \,\varepsilon_{efq}\, \pd^p \pd^q F_{ab}[A]\, .
\end{equation}
This is the double-duality equation defining the field $\phi_{abc}$; notice that this relation is invariant under the gauge transformations
\begin{equation}\label{eq:usualtransfs31}
    \delta \phi_{abc} = 3 \, \pd_{(a} \xi_{bc)}\, , \quad \delta A_a = \pd_a \lambda\, ,
\end{equation}
with a symmetric and \emph{traceful} parameter $\xi_{ab}$. Now, the Maxwell equations for $A_a$ imply that the double trace of $K[\phi]$ vanishes, $\bbar{K}[\phi] \approx 0$, in components
\begin{equation}
    K\indices{_{ab|cd|}^{cd}}[\phi]=0\, .
\end{equation}
This is why the field $\phi_{abc}$ propagates one degree of freedom instead of zero\footnote{The equation of motion of a conventional spin-3 in three dimensions is the \emph{single}-trace equation $\bar{K}[\phi] = 0$, which was shown in \cite{Bekaert:2003az} to be equivalent to the usual, second-order equations 
of motion of Fronsdal \cite{Fronsdal:1978rb}. 
But in three dimensions, the full curvature is determined by its trace, so $\bar{K}[\phi] = 0$ is equivalent to $K[\phi] = 0$ and does not propagate any degree of freedom.}; in fact, this equation and the duality relation together imply that the fields are related algebraically up to a gauge transformation by $\phi_{abc} = \eta_{(ab} A_{c)} + 3 \,\pd_{(a} \Xi_{bc)}$.

All these relations are of third order in derivatives: we will show how they --- or, rather, an adapted version of them containing the extra fields --- come out of a two-derivative Lagragian, which is obtained below by dualising the Maxwell action twice.

\paragraph{Parent action.}

We start from the action of section \ref{sec:3Dscalarmetric} describing 
the off-shell duality between a massless spin-2 field $h_{ab}$ 
and a massless scalar $\phi\,$, through the action \eqref{dualscalarh}. 
Actually, we will use instead the dual action \eqref{dualspin1spin2} that we recall here for convenience: 
\begin{align}
    S[h_{ab},A_a] = \int \!d^3x
    \big[\, &-\tfrac{1}{2}\,\partial_a h_{bc}\,\partial^a h^{bc}\,
    + \tfrac{1}{2}\,\partial_a h^{ab}\,\partial^c h_{cb} 
    + \tfrac{1}{4} F_{ab} F^{ab} 
    + \tfrac{1}{2} \,\varepsilon^{bcd} (\pd^a h_{ab})  F_{cd} \,\big]\;, \label{eq:actionhA}
\end{align}
where $F_{ab} = 2 \pd_{[a} A_{b]}$, in which $\phi$ has been dualised 
to $A_a$ (note that the Fierz-Pauli action does not appear anymore).
This action is invariant under the gauge transformations
\begin{align}
    \delta h_{ab} &= 2 \,\pd_{(a} \epsilon_{b)} \\
    \delta A_a &= \pd_a \lambda + \,\varepsilon_{abc} \,\pd^b \epsilon^c\; .
\end{align}

We now dualize it once more to make a spin-3 appear. 
So, we define the parent action
\begin{align}
    S[G_{a|bc},D_{ab|}{}^{cd},A_a] = \int \!d^3x
    \big[\, &-\tfrac{1}{2}\,G_{a|bc}\,G^{a|bc} + \tfrac{1}{2}\, G_{a|}{}^{ab}\,G^{c|}{}_{cb} + \tfrac{1}{4}\, F_{ab} F^{ab} 
    + \tfrac{1}{2} \,G\indices{^{a|}_{ab}}\, \varepsilon^{bcd} F_{cd}
    \nonumber \\
    &-\tfrac{1}{2} \,\varepsilon^{bcd} G\indices{_{b|a}^a} F_{cd} + G\indices{^{d|}_{bc}}\,\partial^{a}D\indices{_{ad|}^{bc}}\,\big]\;,
    \label{eq:parent21}
\end{align}
with gauge invariances
\begin{align}
\delta A_a &= \partial_a \lambda 
+ \varepsilon_{abc} \,\pd^b \epsilon^c\;, 
\nonumber \\
\delta G_{a|bc} &= 2 \,\pd_a \pd_{(b} \epsilon_{c)} \;, 
\nonumber \\
\delta D_{ab|}{}^{cd} &= 
\varepsilon_{abe}\,\partial^e \upsilon^{cd}
+2 \, \eta^{cd} \pd_{[a} \epsilon_{b]}
+ 4\,\delta_{[a}^{(c} \pd_{b]}
    \epsilon^{d)}\;.
\end{align}
As before, solving the equation of motion for $D\indices{_{ab|}^{cd}}$ 
gives $G_{a|bc} = \pd_a h_{bc}\,$, and we recover the original action 
\eqref{eq:actionhA}. 
Note that we are therefore free to add the term 
$-\tfrac{1}{2} \varepsilon^{bcd} G\indices{_{b|a}^a} F_{cd}$ to the action, 
since this replacement then gives a total derivative\footnote{This term could 
of course come with an arbitrary coefficient. Here, we chose 
$-\tfrac{1}{2}$ in order to simplify some intermediary formulas below; 
however, we checked that keeping it arbitrary does not alter the 
final dual action.}. 
On the other hand, solving the equation of motion of $G$ will give an action 
depending on $D$ and the vector $A\,$; decomposing $D$ following section 
\ref{sec:3Dtop} will make a spin-3 field appear.

The equation of motion for $G$ coming from \eqref{eq:parent21} is
\begin{align}
    0 &= - G_{a|bc} 
    + \eta_{a(b} G\indices{_{c)|d}^d}
   - \pd_d D\indices{_a^{d|}_{bc}} 
   + \tfrac{1}{2} \,\eta_{a(b} \varepsilon_{c)pq} F^{pq} -\tfrac{1}{2}\, \eta_{bc} \,\varepsilon_{apq}\, F^{pq}
\end{align}
and can be solved algebraically for $G_{abc}\,$:
\begin{equation}
    G_{a|bc} = - \pd_d D\indices{_a^{d|}_{bc}} 
    + \eta_{a(b|} \pd_d D\indices{^{ed}_{|c)e}} 
   -\tfrac{1}{2}\, \eta_{bc} \varepsilon_{apq} F^{pq}.
\end{equation}
Substituting this in \eqref{eq:parent21} yields the action
\begin{align}
    S[D,A] = \int \!d^3x
    \big[\, & -\tfrac{1}{4} F_{ab}F^{ab} 
    -\tfrac{1}{2}\,\partial_b D^{ba|c}{}_c\,\varepsilon_{apq}\,F^{pq}
    + \tfrac{1}{2}\,\partial^a D_{ab|cd}\,\partial_e D^{eb|cd} 
    - \tfrac{1}{2}\,\partial^b D_{ba|c}{}^a\,\partial^e D_{ed|}{}^{cd} \big]\;,
    \label{eq:dualAD}
\end{align}
which is invariant under
\begin{align}
\delta A_a &= \partial_a \lambda 
+ \varepsilon_{abc} \,\pd^b \epsilon^c\;, 
\nonumber \\
\delta D_{ab|}{}^{cd} &= 
\varepsilon_{abe}\,\partial^e\upsilon^{cd}
+2 \, \eta^{cd} \pd_{[a} \epsilon_{b]}
+ 4\,\delta^{(c}{}_{[a} \pd_{b]}
    \epsilon^{d)}\;,
\end{align}
as expected. 
We define the three-index tensor
\begin{equation}
    \widetilde{D}^{a| ij} := - \frac{1}{2} \varepsilon^{abc} 
    D_{bc|}{}^{ij} \qquad \Leftrightarrow \qquad D_{ab|}{}^{ij} 
    = \varepsilon_{abc} \widetilde{D}^{c|ij}
\end{equation}
as before. Its gauge transformation is 
\begin{equation}
    \delta \widetilde{D}^{a| ij} = \pd^a \upsilon^{ij} 
    + 2\, \varepsilon^{ab(i}\pd_b \epsilon^{j)} 
    - \eta^{ij} 
    \varepsilon^{abc}\, \pd_b \epsilon_c\, .
\end{equation}
\paragraph{Field redefinitions.}

We now perform the following field redefinitions:
\begin{itemize}
    \item We first define the vector field $U_a$ such that
\begin{align}
    U_a &:= \widetilde{D}_{a|b}{}^b +A_a\;,
    \nonumber \\
    \delta U_a &= \partial_a\upsilon +\partial_a\lambda\;,
    \qquad \upsilon:=\eta^{ab}\,\upsilon_{ab}\;.
    \label{gaugetransfoU}
\end{align}
\item 
We then define the totally symmetric tensor 
\begin{align}
    \phi_{abc}:=&\widetilde{D}_{(a|bc)} 
    - \tfrac{1}{3}\,\eta_{(ab}\,\widetilde{D}_{c)|i}{}^i
    +\tfrac{2}{3} \,\eta_{(ab}\,A_{c)} \;,
    \label{phi}
\end{align}
It transforms according to 
\begin{align}
    \delta\phi_{abc} = \partial_{(a}\widehat{\upsilon}_{bc)} 
    + \tfrac{2}{3}\,\eta_{(ab}\,\partial_{c)}\lambda
    \;,
    \label{gaugeavrphi}
\end{align}
where $\widehat{\upsilon}_{ab}:=\upsilon_{ab}-\frac{1}{3}\,\eta_{ab}\,\upsilon\;$ 
is the traceless part of $\upsilon_{ab}\,$.
Note that the trace 
$\phi_a = \tfrac{10}{9}\,A_a - \tfrac{2}{9}\,\widetilde{D}_{a|}{}^b{}_b 
+ \tfrac{2}{3}\,\widetilde{D}^{b|}{}_{ab}\,$, together with 
$U_a$ and $A_a\,$, gives access to the trace $\widetilde{D}^{b|}{}_{ab}\,$,  
while the other trace $\widetilde{D}_{a|}{}^b{}_b$ is obtainable from 
$U_a$ and $A_a\,$ alone.
\item One then defines the symmetric and traceless tensor
\begin{align}
    f_{ab}:=\varepsilon_{mn(a}\,\widetilde{D}^{m|n}{}_{b)}\;
\end{align}
that transforms like 
\begin{align}
    \delta f_{ab} = 3\,\partial_{(a}\epsilon_{b)} 
    - \eta_{ab}\,\partial^c \epsilon_c + \varepsilon_{pq(a}\,
    \partial^p\widehat{\upsilon}^q{}_{b)}\;.
\end{align}
\item Finally, in order to complete the field spectrum, 
we have to count the gauge field $A_a$ that transforms like
\begin{align}
\delta A_a &= \partial_a \lambda + \varepsilon_{abc} \,\pd^b \epsilon^c\;.
\end{align}
\end{itemize}
The inverse field redefinition is
\begin{align}
    \widetilde{D}_{a|bc} = \phi_{abc} 
        - \tfrac{2}{3}\,\varepsilon_{a(b}{}^p\,f_{c)p}
        +\tfrac{1}{3}\,\eta_{bc}\,U_a - \tfrac{1}{2}\,\eta_{bc}\,\phi_a 
    + \tfrac{1}{2}\,\eta_{a(b}\,\phi_{c)} 
    - \eta_{a(b}\,A_{c)}  \;.  
    \label{dtildedecompo}
\end{align}

\paragraph{Dual action.}

Via the above decomposition \eqref{dtildedecompo}, 
the action \eqref{eq:dualAD} can now be expressed in terms of the fields 
$\{\phi_{abc},f_{ab},A_a,U_a\}$. In the gauge transformations, the only place where the trace $\upsilon^a{}_a$ appears 
is in the transformation law for $U_a\,$, 
meaning that this field appears in the action 
(modulo boundary terms) only through its field strength. Indeed, the action depends on $U_a$ only through the following terms:
\begin{align}
    \cL_U = -\tfrac{1}{18}\,F^{ab}(U)F_{ab}(U)
    - \tfrac{1}{4}\,F^{ab}(U)\big[ F_{ab}(A)+\partial_{[a}\phi_{b]} 
    +\tfrac{2}{3}\,\varepsilon_{abc}\,\partial^d f_{d}{}^c  \big]\;.
\end{align}
This enables us to dualise $U_a$ into a scalar $\sigma$. So, we replace 
$F^{ab}(U)$ everywhere by an independent field $F^{ab}$ and add 
to the resulting Lagrangian the term 
$\varepsilon_{abc}\,F^{ab}\partial^c\sigma\,$.
Variation of the resulting parent Lagrangian with respect to 
$\sigma$ reproduces the original Lagrangian, while $F^{ab}$ is an 
auxiliary field. Substituting its on-shell expression 
in terms of the other fields,
\begin{align}
    F_{ab} \approx 9\,\varepsilon_{abc}\,\partial^c\sigma 
    -\tfrac{9}{4}\,\left(F_{ab}(A)+\partial_{[a}\phi_{b]}
    +\tfrac{2}{3}\,\varepsilon_{abc}\,\partial_d f^{cd}\right)\;
    \label{Fuauxi}
\end{align}
into the parent Lagrangian gives a dual action 
$S[\phi_{abc},f_{ab},A_a,\sigma]\,$ invariant under the 
following gauge transformations 
\begin{align}
    \delta \phi_{abc} &=  \partial_{(a}\widehat{\upsilon}_{bc)} 
                + \tfrac{2}{3}\,\eta_{(ab}\,\partial_{c)}\lambda\;,
    \label{gaugatransfophi}\\
   \delta f_{ab} &= 3\,\partial_{(a}\epsilon_{b)} 
    - \eta_{ab}\,\partial^c \epsilon_c + \varepsilon_{pq(a}\,
    \partial^p\widehat{\upsilon}^q{}_{b)}\;,
    \label{gaugatransfof}\\
   \delta A_a & = \partial_a \lambda + \varepsilon_{abc} \,\pd^b \epsilon^c\;,
    \label{gaugatransfoA}\\
    \delta \sigma &= \tfrac{1}{3}\,\partial_a \epsilon^a\;.
    \label{gaugatransfosigma}
\end{align}
As expected, since the only field transforming with the trace 
$\upsilon^a{}_a$ was the vector $U_a$ that we dualised into 
the scalar $\sigma\,$, after 
dualisation no field transforms with the trace $\upsilon^a{}_a\,$.
Notice also that the quantity on the right-hand side of 
\eqref{Fuauxi}, made out of first derivative of the 
various fields, is gauge invariant since the left-hand side is 
gauge invariant. Upon defining 
$\xi_{ab}:=\widehat{\upsilon}_{ab}+\tfrac{2}{3}\,\eta_{ab}\,\lambda\,$, 
hence $\lambda = \tfrac{1}{2}\,\xi\,$, 
the gauge transformation laws \eqref{gaugatransfophi}-\eqref{gaugatransfoA}
are \emph{exactly} those for the topological gauge system in \eqref{GT1}-\eqref{Utransfo}.
The difference is that, in the present case, we have the extra scalar field 
$\sigma\,$ in the game, with its transformation law \eqref{gaugatransfosigma}.

\paragraph{Analysis of the degrees of freedom.}

We now explicitly prove that the theory under consideration correctly describes 
one degree of freedom per spacetime point, i.e., is equivalent on-shell to the 
theory for a scalar field or its dual vector field in 3D.

For this, one starts from (the dual of) the gauge-invariant quantity obtainable from \eqref{Fuauxi},
\begin{align}
    I^c := \partial^c\sigma - \tfrac{1}{6}\,\partial_{b}f^{bc} 
            +\tfrac{1}{8}\,\varepsilon^{abc}\,\big[ F_{ab}(A) + \partial_a\phi_b\big]\;.
            \label{inva2}
\end{align}
It is easy to see that
\begin{align}
\partial^a I_a\approx 0    \;,
\end{align}
by virtue of the field equation for the scalar field $\sigma\,$.
The antisymmetrised derivatives $\partial_{[a}I_{b]}$ kills the scalar field 
and must therefore vanish on-shell, since it is a gauge-invariant 
quantity independent of the scalar field. Indeed, the study of the topological 
system in Section \ref{sec:3Dtop} showed that all the gauge-invariant 
quantities built out of the fields $\{\phi_{abc},f_{ab},U_a\}$ vanish on-shell (the appearance of $\sigma$ in their equations of motion does not upset the counting argument). As a check, one readily verifies that, as expected,  
\begin{align}\label{eq:divI}
    \varepsilon^{abc}\,\partial_b I_c \equiv -\tfrac{1}{9}\,\left(
    {\cal E}_{(A)}^{a} + \tfrac{2}{3}\,{\cal E}_{(\phi)}^{\;ab}{}_b  
    \right)\approx 0\;,
\end{align}
where ${\cal E}_{(A)}^{a}$ and ${\cal E}_{(\phi)}^{abc}$ denote the 
left-hand sides of the field equations for $A_a$ and $\phi_{abc}\,$, 
respectively. Similarly, ${\cal E}_{(f)}^{ab}$ and ${\cal E}_{(\sigma)}$
denote the left-hand sides of the field equations for $f_{ab}$ and 
$\sigma\,$.
We then have the symmetrised derivative $\partial_{(a}I_{b)}\,$,  
whose traceless part $\widehat{I}_{ab}$ does not vanish on shell 
and that constitutes part of the set of gauge invariant quantities 
that are non-vanishing on shell. 
At the next order in the derivatives of $I_a\,$, we decompose 
$\partial_{a}\partial_{b}I_{c}$ into irreducible representations of 
$so(3)$ and find that only the spin-3 representation, i.e., the traceless
component $\widehat{I}_{abc}$ of the symmetrised derivatives 
$\partial_{(a}\partial_{b}I_{c)}$ is not vanishing on shell. In order to 
see this, we used that $\partial^a I_a\approx 0 \approx \partial_{[a}I_{b]}$ 
and that $\Box I_a\approx 0\,$. It is indeed direct to see that 
\begin{align}
\Box I_a \equiv -\tfrac{1}{9}\,\varepsilon_{abc}\,\left(
\partial^b {\cal E}_{(A)}^c + \tfrac{2}{3}\,\partial^b {\cal E}_{(\phi)}^{\;cd}{}_d
\right) +\tfrac{1}{18}\,\partial_a {\cal E}_{(\sigma)} \approx 0\;.   
\label{boxI}
\end{align}
From then on, it is straightforward to conclude that the only gauge-invariant quantities 
that are not vanishing on shell are the traceless parts 
$\widehat{I}_{a_1 a_2\ldots a_{n+1}}$ of the 
symmetrised derivatives $\partial_{(a_1}\ldots \partial_{a_n}I_{a_{n+1})}\,$:
\begin{align}
    {\cal T} = \{\widehat{I}_{a_1\ldots a_{n+1}}\;, \quad n\in\mathbb{N} \}\;,\qquad 
    \widehat{I}_a := I_a\;.
\end{align}
This infinite set of gauge-invariant tensors that are not vanishing on shell 
indeed characterises a theory for a propagating masseless scalar field, 
see e.g. the review \cite{Bekaert:2005vh}.

\paragraph{Equations of motion and duality relations.} From the expressions \eqref{gaugatransfof} and \eqref{gaugatransfosigma}, 
one observes that it is possible to absorb the traceless 
field $f_{ab}$ and the scalar field $\sigma$ into a traceful 
field $h_{ab}\,$ via the redefinition
\begin{align}
    h_{ab}:= \tfrac{2}{3}\,f_{ab} + 2\,\eta_{ab}\,\sigma\;,
    \qquad \sigma = \tfrac{1}{6}\,h^a{}_a\;.
\end{align}
The resulting action $S[\phi_{abc},h_{ab},A_a]$ is then invariant under 
\begin{align}
 \delta \phi_{abc} &=  \partial_{(a}{\xi}_{bc)} \;,
    \label{gaugetransfophi}\\
   \delta h_{ab} &= 2\,\partial_{(a}\epsilon_{b)} 
     + \tfrac{2}{3}\,\varepsilon_{pq(a}\,\partial^p\xi^q{}_{b)}\;,
    \label{gaugetransfoh}\\
   \delta A_a & = \tfrac{1}{2}\,\partial_a \xi + \varepsilon_{abc} \,\pd^b \epsilon^c\;,
    \label{gaugetransfoA}
\end{align}
where the traceless symmetric tensor $\widehat{\upsilon}_{ab}$ and the scalar 
$\lambda$ have been combined into a traceful symmetric parameter $\xi_{ab}$. 
As in section \ref{sec:3Dtop} and unlike the two-column case, the gauge 
transformations cannot be disentangled. If one wishes, upon 
redefining the spin-3 field $\phi_{abc}$ 
by adding pure-trace contributions involving the spin-1 field $A_a\,$, one can make appear the symmetric gradient 
$\partial_{(a}\widehat{\xi}_{bc)}$ of the \emph{traceless} parameter 
$\widehat{\xi}_{ab}$
in the gauge transformation law of the newly defined spin-3 field 
$\varphi_{abc}\,$, thereby producing the correct gauge-transformation 
law for a massless spin-3 gauge field. This, however, is done at the expense 
of extra terms in the gauge transformation law, terms that depend on the 
spin-2 gauge parameter $\epsilon_a\,$, as in section \ref{sec:3Dtop}.

Now, we show how we recover the equations of motion for the various fields. Notice first that the spin-3 field strength
\begin{align}
    K_{ab|cd|ef}[\phi] := 8 \,\pd_{[a} \pd_{[c} \pd_{[e} \phi_{f]d]b]}\;
\end{align}
is a gauge invariant quantity of the theory. Therefore, on-shell, it must either vanish or be proportional to a derivative of the invariant $I_a\,$ defined in \eqref{inva2} that is not 
vanishing on shell. It turns out that it is  
the second possibility that is actually realised:
\begin{equation}\label{eq:dualityK}
    \cK_{abc} \approx
-6\, \partial_{(a}\partial_{b}I_{c)}
\end{equation}
where
\begin{equation}
    \cK^{mnp}[\phi] := \tfrac{1}{8} \varepsilon^{mab}\varepsilon^{ncd}
    \varepsilon^{pef}\, K_{ab|cd|ef}[\phi]\, .
\end{equation}
Therefore, using the trace identity $\bbar{K}_{ab} = 2\, \varepsilon_{abc} \bar{\cK}^c$ and the result --- obtained above \eqref{boxI} --- that the trace of 
$\partial_{(\alpha}\partial_{\beta}I_{\gamma)}$ vanishes on shell, we find indeed the required double-trace equation for the spin three field,
\begin{equation}
    \bbar{K}_{ab}[\phi] \approx 0\, .
\end{equation}
For the spin one field, from \eqref{eq:divI} we have
\begin{equation}
    \pd_a \widetilde{F}^{ab} \approx 0\, ,
\end{equation}
i.e. the usual equations but where 
\begin{equation}
    \widetilde{F}_{ab} := - 4 \,\varepsilon_{abc} I^c = F_{ab}[A] + \pd_{[a} \phi_{b]c}{}^c + \varepsilon_{abc} \left( \pd_d h^{cd} - \pd^c h^d{}_d \right)
\end{equation}
is the field strength $F_{ab}[A]$, with correction terms involving the other fields in order to make it gauge-invariant. Similarly for the spin $2$ field, the gauge-invariant tensor that corrects the linearised Riemann tensor is
\begin{equation}
    \widetilde{R}_{ab|}{}^{cd} = {\cR}_{ab}{}^{cd}[h] -2 \left( \varepsilon_{abm} \pd^{[c} \Psi^{d]m} + \varepsilon^{cdm} \pd_{[a} \Psi_{b]m} \right)\, , \quad \Psi^a{}_b:= \pd_b \phi^{ac}{}_c - \pd^c \phi^{a}{}_{bc}\, .
\end{equation}
On-shell, we have
\begin{equation}\label{eq:dualityG}
    \widetilde{G}_{ab}\approx -{7}\,\partial_{(a}I_{b)}\, , \quad \widetilde{G}_{ab} := \frac{1}{4} \varepsilon_{amn} \varepsilon_{bpq} \widetilde{R}^{mn|pq}\, .
\end{equation}
On account of $\pd^a I_a \approx 0$ and the trace identity $\widetilde{G}^a{}_a = -\tfrac{1}{2} \widetilde{R}^{ab|}{}_{ab}$, this implies the correct double-trace equation
\begin{equation}
    \widetilde{R}^{ab|}{}_{ab} \approx 0
\end{equation}
for a propagating field $h_{ab}$ in $D=3$ (as explained in section \ref{sec:3Dscalarmetric}), suitably modified to make it gauge-invariant.

We also recover the duality relations between those three fields: they are exactly equations \eqref{eq:dualityK} and \eqref{eq:dualityG}. Indeed, with some $\varepsilon$-tensor manipulations, \eqref{eq:dualityK} can be written as
\begin{equation}\label{eq:dualityKF}
    K_{ab|cd|ef}[\phi] \approx -\tfrac{2}{3}\, \varepsilon_{cdp} \,\varepsilon_{efq}\,\pd^p \pd^q \widetilde{F}_{ab}\, .
\end{equation}
This is the version of the duality equation \eqref{eq:dualityKFnaive} that is invariant under \eqref{gaugatransfophi}---\eqref{gaugetransfoA}. Similarly, equation \eqref{eq:dualityG} is the same as
\begin{equation}\label{eq:dualityR}
    \widetilde{R}^{ab|}{}_{cd} \approx \tfrac{7}{4}\, \varepsilon_{cdm} \pd^m \widetilde{F}^{ab}.
\end{equation}
which is the invariant version of the duality relation between $h_{ab}$ and $A_a$, obtained by dualising $A_a$ on an empty column. To close the loop, one can also use \eqref{eq:dualityR} in \eqref{eq:dualityKF} to get the expected duality relation between $\phi_{abc}$ and $h_{ab}$,
\begin{equation}
    K_{ab|cd|ef}[\phi] \approx -\tfrac{8}{21} \,\varepsilon_{efq}\,\pd^q \widetilde{R}_{ab|cd}\, .
\end{equation}
So, all the duality relations and equations of motion come out of the action. The price to pay is that the gauge transformations are not the usual, independent ones for each field: therefore, one actually gets a suitably gauge-invariant version of those equations.

\section{Conclusions}

In this paper, we have revisited the problem of writing down actions for 
higher (or exotic) dual gauge fields, whose equations of motion take the 
form of multiple traces of the gauge-invariant field strength. 
Examples of such fields include the double-dual graviton of 
\cite{Hull:2001iu}, or the exotic duals of supergravity $p$-form 
fields with extra sets of $D-2$ antisymmetric indices \cite{Riccioni:2006az,West:2004kb,Cook:2009ri,Bergshoeff:2011se,Chatzistavrakidis:2013jqa,West:2018lfn,Fernandez-Melgarejo:2018yxq}. 
Actions for those fields were first derived in 
\cite{Boulanger:2012df,Boulanger:2012mq,Boulanger:2015mka} from those for the 
original (non-dualised) fields using the method of parent actions, therefore 
realising the duality off-shell and in manifestly covariant form. 
A common feature of these actions is that they include a number of extra fields 
that cannot be eliminated from the action. We started from those actions and 
performed some change of variables/dualisations of the additional fields, 
clarifying their role and the counting of degrees of freedom. 
In particular, for the higher spin case where the Young diagram of the field 
contains more than two columns, these action produce a version of the 
duality relations corrected by the additional fields.

Of course, these considerations apply only to the linear theory; 
extending them to a putative interacting theory is still an outstanding challenge. 
In this context, the coupling to external sources is of particular interest and 
could give some insight into the exotic branes of string theory which are charged 
under the mixed symmetry potentials considered here \cite{Bergshoeff:2011se,Chatzistavrakidis:2013jqa}. 
A precise link between these branes and the additional fields that appear 
from the off-shell dualisation is however still missing (see also 
\cite{Bergshoeff:2016gub} for comments on this point in the context of Double 
Field Theory). Contact should also be made with the work of \cite{Bunster:2013era}, 
in which classical sources for mixed symmetry fields are defined using the 
idea of a `brane within a brane'.

In 3D, one can in principle keep going, performing more and more  
dualisations. In the topological case, 
it is natural to conjecture that this off-shell dualisation procedure 
will lead to an action with a spectrum of gauge fields given by  
$\{h_{a_1a_2},\varphi_{a_1a_2a_3},\ldots, \varphi_{a_1\ldots a_s}\,\}$, 
while in the non-topological case, one will obtain 
an action featuring the gauge fields 
$\{A_a,h_{a_1a_2},\varphi_{a_1a_2a_3},\ldots, \varphi_{a_1\ldots a_s}\,\}$.
In both cases, the gauge transformations of the all the fields 
are intermingled in the sense that all the gauge parameters that can 
appear in a transformation law of a field will effectively appear.
In both cases, one may speculate that the action could be presented 
in a Chern-Simons form, which would facilitate the introduction of 
interactions \cite{inprogress}. 
Of course, it would also be very interesting to take the limit where 
the highest spin of the spectrum goes to infinity. 

\paragraph{Acknowledgements.}

We would like to thank Andrea Campoleoni, Chris Hull and Zhenya Skvortsov 
for useful conversations. 
We performed or checked several computations with the package xTras 
\cite{Nutma:2013zea} of the suite of Mathematica packages xAct.
The work of NB is partially supported by an F.R.S.-FNRS PDR grant 
``Fundamental issues in extended gravity'' No T.0022.19.
The work of VL was supported by the European Research Council (ERC) under the 
European Union’s Horizon 2020 research and innovation programme, grant agreements No. 740209 (eQG ERC–2016–ADG) and No. 724659  (MassiveCosmo ERC–2016–COG).


\begin{thebibliography}{10}

\bibitem{Hull:2000zn}
C.~Hull, ``{Strongly coupled gravity and duality},''
  \href{http://dx.doi.org/10.1016/S0550-3213(00)00323-0}{{\em Nucl. Phys. B}
  {\bfseries 583} (2000) 237--259},
  \href{http://arxiv.org/abs/hep-th/0004195}{{\ttfamily arXiv:hep-th/0004195}}.

\bibitem{Hull:2000rr}
C.~Hull, ``{Symmetries and compactifications of (4,0) conformal gravity},''
  \href{http://dx.doi.org/10.1088/1126-6708/2000/12/007}{{\em JHEP} {\bfseries
  12} (2000) 007}, \href{http://arxiv.org/abs/hep-th/0011215}{{\ttfamily
  arXiv:hep-th/0011215}}.

\bibitem{Hull:2001iu}
C.~Hull, ``{Duality in gravity and higher spin gauge fields},''
  \href{http://dx.doi.org/10.1088/1126-6708/2001/09/027}{{\em JHEP} {\bfseries
  09} (2001) 027}, \href{http://arxiv.org/abs/hep-th/0107149}{{\ttfamily
  arXiv:hep-th/0107149}}.

\bibitem{West:2001as}
P.~C. West, ``{E(11) and M theory},''
  \href{http://dx.doi.org/10.1088/0264-9381/18/21/305}{{\em Class. Quant.
  Grav.} {\bfseries 18} (2001) 4443--4460},
  \href{http://arxiv.org/abs/hep-th/0104081}{{\ttfamily arXiv:hep-th/0104081}}.

\bibitem{West:2002jj}
P.~C. West, ``{Very extended E(8) and A(8) at low levels, gravity and
  supergravity},'' \href{http://dx.doi.org/10.1088/0264-9381/20/11/328}{{\em
  Class. Quant. Grav.} {\bfseries 20} (2003) 2393--2406},
  \href{http://arxiv.org/abs/hep-th/0212291}{{\ttfamily arXiv:hep-th/0212291}}.

\bibitem{Boulanger:2003vs}
N.~Boulanger, S.~Cnockaert, and M.~Henneaux, ``{A note on spin s duality},''
  \href{http://dx.doi.org/10.1088/1126-6708/2003/06/060}{{\em JHEP} {\bfseries
  06} (2003) 060}, \href{http://arxiv.org/abs/hep-th/0306023}{{\ttfamily
  arXiv:hep-th/0306023}}.

\bibitem{Curtright:1980yk}
T.~Curtright, ``{Generalized Gauge Fields},''
\href{http://dx.doi.org/10.1016/0370-2693(85)91235-3}{{\em Phys. Lett.}
  {\bfseries 165B} (1985) 304--308}.

\bibitem{Aulakh:1986cb}
C.~S. Aulakh, I.~G. Koh, and S.~Ouvry, ``{Higher Spin Fields With Mixed
  Symmetry},''
\href{http://dx.doi.org/10.1016/0370-2693(86)90518-6}{{\em Phys. Lett.}
  {\bfseries B173} (1986) 284--288}.

\bibitem{Riccioni:2006az}
F.~Riccioni and P.~C. West, ``{Dual fields and E(11)},''
  \href{http://dx.doi.org/10.1016/j.physletb.2006.12.050}{{\em Phys. Lett. B}
  {\bfseries 645} (2007) 286--292},
  \href{http://arxiv.org/abs/hep-th/0612001}{{\ttfamily arXiv:hep-th/0612001}}.

\bibitem{Bekaert:2002dt}
X.~Bekaert and N.~Boulanger, ``{Tensor gauge fields in arbitrary
  representations of GL(D,R): Duality and Poincare lemma},''
  \href{http://dx.doi.org/10.1007/s00220-003-0995-1}{{\em Commun. Math. Phys.}
  {\bfseries 245} (2004) 27--67},
  \href{http://arxiv.org/abs/hep-th/0208058}{{\ttfamily arXiv:hep-th/0208058}}.

\bibitem{deMedeiros:2002qpr}
P.~de~Medeiros and C.~Hull, ``{Exotic tensor gauge theory and duality},''
  \href{http://dx.doi.org/10.1007/s00220-003-0810-z}{{\em Commun. Math. Phys.}
  {\bfseries 235} (2003) 255--273},
  \href{http://arxiv.org/abs/hep-th/0208155}{{\ttfamily arXiv:hep-th/0208155}}.

\bibitem{Bekaert:2003az}
X.~Bekaert and N.~Boulanger, ``{On geometric equations and duality for free
  higher spins},'' \href{http://dx.doi.org/10.1016/S0370-2693(03)00409-X}{{\em
  Phys. Lett. B} {\bfseries 561} (2003) 183--190},
  \href{http://arxiv.org/abs/hep-th/0301243}{{\ttfamily arXiv:hep-th/0301243}}.

\bibitem{Boulanger:2012df}
N.~Boulanger, P.~P. Cook, and D.~Ponomarev, ``{Off-Shell Hodge Dualities in
  Linearised Gravity and E11},''
  \href{http://dx.doi.org/10.1007/JHEP09(2012)089}{{\em JHEP} {\bfseries 09}
  (2012) 089},
\href{http://arxiv.org/abs/1205.2277}{{\ttfamily arXiv:1205.2277 [hep-th]}}.

\bibitem{Boulanger:2015mka}
N.~Boulanger, P.~Sundell, and P.~West, ``{Gauge fields and infinite chains of
  dualities},'' \href{http://dx.doi.org/10.1007/JHEP09(2015)192}{{\em JHEP}
  {\bfseries 09} (2015) 192},
\href{http://arxiv.org/abs/1502.07909}{{\ttfamily arXiv:1502.07909 [hep-th]}}.

\bibitem{West:2004kb}
P.~C. West, ``{E(11) origin of brane charges and U-duality multiplets},''
  \href{http://dx.doi.org/10.1088/1126-6708/2004/08/052}{{\em JHEP} {\bfseries
  08} (2004) 052}, \href{http://arxiv.org/abs/hep-th/0406150}{{\ttfamily
  arXiv:hep-th/0406150}}.

\bibitem{Cook:2009ri}
P.~P. Cook, ``{Exotic E(11) branes as composite gravitational solutions},''
  \href{http://dx.doi.org/10.1088/0264-9381/26/23/235023}{{\em Class. Quant.
  Grav.} {\bfseries 26} (2009) 235023},
  \href{http://arxiv.org/abs/0908.0485}{{\ttfamily arXiv:0908.0485 [hep-th]}}.

\bibitem{Bergshoeff:2011se}
E.~A. Bergshoeff, T.~Ortin, and F.~Riccioni, ``{Defect Branes},''
  \href{http://dx.doi.org/10.1016/j.nuclphysb.2011.10.037}{{\em Nucl. Phys. B}
  {\bfseries 856} (2012) 210--227},
  \href{http://arxiv.org/abs/1109.4484}{{\ttfamily arXiv:1109.4484 [hep-th]}}.

\bibitem{Chatzistavrakidis:2013jqa}
A.~Chatzistavrakidis, F.~F. Gautason, G.~Moutsopoulos, and M.~Zagermann,
  ``{Effective actions of nongeometric five-branes},''
  \href{http://dx.doi.org/10.1103/PhysRevD.89.066004}{{\em Phys. Rev. D}
  {\bfseries 89} no.~6, (2014) 066004},
  \href{http://arxiv.org/abs/1309.2653}{{\ttfamily arXiv:1309.2653 [hep-th]}}.

\bibitem{West:2018lfn}
P.~West, ``{E$_{11}$, Brane Dynamics and Duality Symmetries},''
  \href{http://dx.doi.org/10.1142/S0217751X1850080X}{{\em Int. J. Mod. Phys. A}
  {\bfseries 33} no.~13, (2018) 1850080},
  \href{http://arxiv.org/abs/1801.00669}{{\ttfamily arXiv:1801.00669
  [hep-th]}}.

\bibitem{Fernandez-Melgarejo:2018yxq}
J.~J. Fern\'andez-Melgarejo, T.~Kimura, and Y.~Sakatani, ``{Weaving the Exotic
  Web},'' \href{http://dx.doi.org/10.1007/JHEP09(2018)072}{{\em JHEP}
  {\bfseries 09} (2018) 072}, \href{http://arxiv.org/abs/1805.12117}{{\ttfamily
  arXiv:1805.12117 [hep-th]}}.

\bibitem{Henneaux:2019zod}
M.~Henneaux, V.~Lekeu, and A.~Leonard, ``{A note on the double dual
  graviton},'' \href{http://dx.doi.org/10.1088/1751-8121/ab56ed}{{\em J. Phys.}
  {\bfseries A53} no.~1, (2020) 014002},
\href{http://arxiv.org/abs/1909.12706}{{\ttfamily arXiv:1909.12706 [hep-th]}}.

\bibitem{Chatzistavrakidis:2019bxo}
A.~Chatzistavrakidis and G.~Karagiannis, ``{Relation between standard and
  exotic duals of differential forms},''
  \href{http://dx.doi.org/10.1103/PhysRevD.100.121902}{{\em Phys. Rev. D}
  {\bfseries 100} no.~12, (2019) 121902},
  \href{http://arxiv.org/abs/1911.00419}{{\ttfamily arXiv:1911.00419
  [hep-th]}}.

\bibitem{Chatzistavrakidis:2020kpx}
A.~Chatzistavrakidis, G.~Karagiannis, and A.~Ranjbar, ``{Duality and higher
  Buscher rules in p-form gauge theory and linearized gravity},''
  \href{http://arxiv.org/abs/2012.08220}{{\ttfamily arXiv:2012.08220
  [hep-th]}}.

\bibitem{Boulanger:2012mq}
N.~Boulanger and D.~Ponomarev, ``{Frame-like off-shell dualisation for
  mixed-symmetry gauge fields},''
  \href{http://dx.doi.org/10.1088/1751-8113/46/21/214014}{{\em J. Phys. A}
  {\bfseries 46} (2013) 214014},
  \href{http://arxiv.org/abs/1206.2052}{{\ttfamily arXiv:1206.2052 [hep-th]}}.

\bibitem{Bergshoeff:2016ncb}
E.~A. Bergshoeff, O.~Hohm, V.~A. Penas, and F.~Riccioni, ``{Dual Double Field
  Theory},'' \href{http://dx.doi.org/10.1007/JHEP06(2016)026}{{\em JHEP}
  {\bfseries 06} (2016) 026}, \href{http://arxiv.org/abs/1603.07380}{{\ttfamily
  arXiv:1603.07380 [hep-th]}}.

\bibitem{Bergshoeff:2016gub}
E.~A. Bergshoeff, O.~Hohm, and F.~Riccioni, ``{Exotic Dual of Type II Double
  Field Theory},'' \href{http://dx.doi.org/10.1016/j.physletb.2017.01.081}{{\em
  Phys. Lett. B} {\bfseries 767} (2017) 374--379},
  \href{http://arxiv.org/abs/1612.02691}{{\ttfamily arXiv:1612.02691
  [hep-th]}}.

\bibitem{Chatzistavrakidis:2019len}
A.~Chatzistavrakidis, G.~Karagiannis, and P.~Schupp, ``{A unified approach to
  standard and exotic dualizations through graded geometry},''
  \href{http://dx.doi.org/10.1007/s00220-020-03728-x}{{\em Commun. Math. Phys.}
  {\bfseries 378} no.~2, (2020) 1157--1201},
  \href{http://arxiv.org/abs/1908.11663}{{\ttfamily arXiv:1908.11663
  [hep-th]}}.

\bibitem{Burdik:2000kj}
C.~Burdik, A.~Pashnev, and M.~Tsulaia, ``{The Lagrangian description of
  representations of the Poincare group},''
  \href{http://dx.doi.org/10.1016/S0920-5632(01)01568-7}{{\em Nucl. Phys. B
  Proc. Suppl.} {\bfseries 102} (2001) 285--292},
  \href{http://arxiv.org/abs/hep-th/0103143}{{\ttfamily arXiv:hep-th/0103143}}.

\bibitem{Bekaert:2004dz}
X.~Bekaert, N.~Boulanger, and S.~Cnockaert, ``{No self-interaction for
  two-column massless fields},''
  \href{http://dx.doi.org/10.1063/1.1823032}{{\em J. Math. Phys.} {\bfseries
  46} (2005) 012303}, \href{http://arxiv.org/abs/hep-th/0407102}{{\ttfamily
  arXiv:hep-th/0407102}}.

\bibitem{Grigoriev:2020lzu}
M.~Grigoriev, K.~Mkrtchyan, and E.~Skvortsov, ``{Matter-free higher spin
  gravities in 3D: Partially-massless fields and general structure},''
  \href{http://dx.doi.org/10.1103/PhysRevD.102.066003}{{\em Phys. Rev. D}
  {\bfseries 102} no.~6, (2020) 066003},
  \href{http://arxiv.org/abs/2005.05931}{{\ttfamily arXiv:2005.05931
  [hep-th]}}.

\bibitem{inprogress}
N.~Boulanger, A.~Campoleoni, V.~Lekeu, and E.~Skvortsov~(in progress).

\bibitem{DuboisViolette:2001jk}
M.~Dubois-Violette and M.~Henneaux, ``{Tensor fields of mixed Young symmetry
  type and N complexes},'' \href{http://dx.doi.org/10.1007/s002200200610}{{\em
  Commun. Math. Phys.} {\bfseries 226} (2002) 393--418},
  \href{http://arxiv.org/abs/math/0110088}{{\ttfamily arXiv:math/0110088}}.

\bibitem{Fronsdal:1978rb}
C.~Fronsdal, ``{Massless Fields with Integer Spin},''
  \href{http://dx.doi.org/10.1103/PhysRevD.18.3624}{{\em Phys. Rev. D}
  {\bfseries 18} (1978) 3624}.

\bibitem{Bekaert:2005vh}
X.~Bekaert, S.~Cnockaert, C.~Iazeolla, and M.~Vasiliev, ``{Nonlinear higher
  spin theories in various dimensions},'' in {\em {1st Solvay Workshop on
  Higher Spin Gauge Theories}}, pp.~132--197.
\newblock 2004.
\newblock \href{http://arxiv.org/abs/hep-th/0503128}{{\ttfamily
  arXiv:hep-th/0503128}}.

\bibitem{Bunster:2013era}
C.~Bunster and M.~Henneaux, ``{Sources for Generalized Gauge Fields},''
  \href{http://dx.doi.org/10.1103/PhysRevD.88.085002}{{\em Phys. Rev. D}
  {\bfseries 88} (2013) 085002},
  \href{http://arxiv.org/abs/1308.2866}{{\ttfamily arXiv:1308.2866 [hep-th]}}.

\bibitem{Nutma:2013zea}
T.~Nutma, ``{xTras : A field-theory inspired xAct package for mathematica},''
  \href{http://dx.doi.org/10.1016/j.cpc.2014.02.006}{{\em Comput. Phys.
  Commun.} {\bfseries 185} (2014) 1719--1738},
  \href{http://arxiv.org/abs/1308.3493}{{\ttfamily arXiv:1308.3493 [cs.SC]}}.

\end{thebibliography}

\providecommand{\href}[2]{#2}\begingroup\raggedright\endgroup

\end{document}